\documentclass[twocolumn]{aastex63}

\usepackage{natbib}
\graphicspath{{./}{Figures/}}

\usepackage{hyperref}

\newcommand{\uat}[2]{\href{http://astrothesaurus.org/uat/#2}{#1 (#2)}}

\def\xmm{{\it XMM-Newton}}
\defcitealias{PaperI}{Paper I}
\defcitealias{PaperII}{Paper II}
\defcitealias{PaperIII}{Paper III}
\defcitealias{PaperIV}{Paper IV}
\defcitealias{PaperV}{Paper V}
\defcitealias{PaperVI}{Paper VI}

\shorttitle{GraL VII -- XMM-Newton Observations of Lensed Quasars}
\shortauthors{Connor et al.}

\begin{document}

\title{Gaia GraL: Gaia DR2 Gravitational Lens Systems. VII.\\
XMM-Newton Observations of Lensed Quasars}

\correspondingauthor{Thomas Connor}
\email{thomas.p.connor@jpl.nasa.gov}

\suppressAffiliations
\author[0000-0002-7898-7664]{Thomas Connor}
\altaffiliation{NPP Fellow}
\affiliation{Jet Propulsion Laboratory, California Institute of Technology, 4800 Oak Grove Drive, Pasadena, CA 91109, USA}

\author[0000-0003-2686-9241]{Daniel Stern}
\affiliation{Jet Propulsion Laboratory, California Institute of Technology, 4800 Oak Grove Drive, Pasadena, CA 91109, USA}

\author[0000-0002-2308-6623]{Alberto Krone-Martins}
\affiliation{Donald Bren School of Information and Computer Sciences, University of California, Irvine, Irvine CA 92697, USA}

\author[0000-0002-0603-3087]{S. G. Djorgovski}
\affiliation{Division of Physics, Mathematics, and Astronomy, Caltech, Pasadena, CA 91125}

\author[0000-0002-3168-0139]{Matthew J. Graham}
\affiliation{Division of Physics, Mathematics, and Astronomy, Caltech, Pasadena, CA 91125}

\author[0000-0001-5819-3552]{Dominic J. Walton}
\affiliation{Institute of Astronomy, University of Cambridge, Madingley Road, Cambridge CB3 0HA, UK}

\nocollaboration{6}

\author[0000-0003-2559-408X]{Ludovic Delchambre}
\affiliation{Institut d'Astrophysique et de G\'{e}ophysique, Universit\'{e} de Li\`{e}ge, 19c, All\'{e}e du 6 Ao\^{u}t, B-4000 Li\`{e}ge, Belgium}

\author[0000-0003-4843-8979]{Christine Ducourant}
\affiliation{Laboratoire d'Astrophysique de Bordeaux, Univ. Bordeaux, CNRS, B18N, all{\'e}e Geoffroy Saint-Hilaire, 33615 Pessac, France} 

\author[0000-0002-6806-6626]{Ramachrisna Teixeira}
\affiliation{Instituto de Astronomia, Geofísica e Ciências Atmosféricas, Universidade de São Paulo, Rua do Matão, 1226, Cidade Universitária, 05508-900 São Paulo, SP, Brazil}

\author{Jean-Fran\c{c}ois Le Campion}
\affiliation{Laboratoire d'Astrophysique de Bordeaux, Univ. Bordeaux, CNRS, B18N, all{\'e}e Geoffroy Saint-Hilaire, 33615 Pessac, France}

\author[0000-0002-8760-6157]{Jakob Sebastian den Brok}
\affiliation{Argelander-Institut f\"ur Astronomie, Universit\"at Bonn, Auf dem H\"ugel 71, 53121 Bonn, Germany}

\author[0000-0003-0699-7019]{Dougal Dobie}
\affiliation{Centre for Astrophysics and Supercomputing, Swinburne University of Technology, Hawthorn, Victoria, Australia}
\affiliation{ARC Centre of Excellence for Gravitational Wave Discovery (OzGrav), Hawthorn, Victoria, Australia}

\author[0000-0002-8541-0476]{Laurent Galluccio}
\affiliation{Université Côte d'Azur, Observatoire de la Côte d'Azur, CNRS, Laboratoire Lagrange, Bd de l'Observatoire, CS 34229, 06304 Nice cedex 4, France}

\author[0000-0002-0524-5328]{Priyanka Jalan}
\affiliation{Aryabhatta Research Institute of Observational Sciences (ARIES), Manora Peak, Nainital, 263002 India}

\author[0000-0003-4682-7831]{Sergei A. Klioner}
\affiliation{Lohrmann-Observatorium, Technische Universit\"at Dresden,
01062 Dresden, Germany}

\author[0000-0002-3469-5133]{Jonas Kl\"{u}ter}
\affiliation{Department of Physics and Astronomy, Louisiana State University, Baton Rouge, LA 70803 USA}

\author[0000-0003-2242-0244]{Ashish A. Mahabal}
\affiliation{Division of Physics, Mathematics, and Astronomy, Caltech, Pasadena, CA 91125}
\affiliation{Center for Data Driven Discovery, California Institute of Technology, Pasadena, CA 91125, USA}

\author[0000-0001-5824-1040]{Vibhore Negi}
\affiliation{Aryabhatta Research Institute of Observational Sciences (ARIES), Manora Peak, Nainital, 263002 India}

\author[0000-0001-6809-2536]{Anna Nierenberg}
\affiliation{Department of Physics, University of California Merced, 5200 North Lake Road, Merced, CA 95343, USA}

\author{Quentin Petit}
\affiliation{Laboratoire d'Astrophysique de Bordeaux, Univ. Bordeaux, CNRS, B18N, all{\'e}e Geoffroy Saint-Hilaire, 33615 Pessac, France}

\author[00000-0003-3739-4288]{Sergio Scarano Jr}
\altaffiliation{IAG-USP Fellow}
\affiliation{Departamento de F\'{i}sica – CCET, Universidade Federal de Sergipe, Rod. Marechal Rondon s/n, 49.100-000, Jardim Rosa Elze, S\~{a}o Crist\'{o}v\~{a}o, SE, Brazil}

\author{Eric Slezak}
\affiliation{Universit\'{e} C\^{o}te d'Azur, Observatoire de la C\^{o}te d'Azur, CNRS, Laboratoire Lagrange, Boulevard de l'Observatoire, CS 34229, 06304 Nice, France}

\author[0000-0001-6116-2095]{Dominique Sluse}
\affiliation{Institut d'Astrophysique et de G\'{e}ophysique, Universit\'{e} de Li\`{e}ge, 19c, All\'{e}e du 6 Ao\^{u}t, B-4000 Li\`{e}ge, Belgium}

\author[0000-0002-8052-7763]{Carolina Spíndola-Duarte}
\affiliation{Instituto de Astronomia, Geofísica e Ciências Atmosféricas, Universidade de São Paulo, Rua do Matão, 1226, Cidade Universitária, 05508-900 São Paulo, SP, Brazil}

\author[0000-0002-7005-1976]{Jean Surdej}
\affiliation{Space sciences, Technologies and Astrophysics Research (STAR) Institute, University of Li\`ege, Belgium}
\affiliation{Astronomical Observatory Institute, Faculty of Physics, Adam Mickiewicz University, ul. Sloneczna 36, 60-286 Pozna\'n, Poland}

\author[0000-0002-8365-7619]{Joachim Wambsganss}
\affiliation{Zentrum f\"{u}r Astronomie der Universit\"{a}t Heidelberg, Astronomisches Rechen-Institut, M\"{o}nchhofstr. 12-14, 69120 Heidelberg,\\Germany}

\collaboration{3}{The Gaia GraL Team}

\begin{abstract}

We present \textit{XMM-Newton} X-ray observations of nine confirmed lensed quasars at $1 \lesssim z \lesssim 3$ identified by the \textit{Gaia} Gravitational Lens program. Eight systems are strongly detected, with 0.3--8.0 keV fluxes $F_{0.3-8.0} \gtrsim 5 \times 10^{-14}\ {\rm erg}\ {\rm cm}^{-2}\ {\rm s}^{-1}$. Modeling the X-ray spectra with an absorbed power law, we derive power law photon indices and 2--10 keV luminosities for the eight detected quasars. In addition to presenting sample properties for larger quasar population studies and for use in planning for future caustic crossing events, we also identify three quasars of interest: a quasar that shows evidence of flux variability from previous \textit{ROSAT} observations, the most closely-separated individual lensed sources resolved by \textit{XMM-Newton}, and one of the X-ray brightest quasars known at $z>3$. These sources represent the tip of discovery that will be enabled by \textit{SRG}/eROSITA. 

\end{abstract}


\section{Introduction} 
\begin{deluxetable*}{ccccccl}
\tablecaption{Target properties and \xmm\, observations.}
\label{table:sample}
\tablehead{
\colhead{Target} &
\colhead{R.A.} &
\colhead{Dec.} &
\colhead{$z$} &
\colhead{$\mu$\tablenotemark{a}} &
\colhead{Separation\tablenotemark{b}} &
\colhead{Notes}}
\decimals
\startdata
GraL J0659+1629   & 06:59:04.1 &   +16:29:09 & 3.083 & 37.6  & 6\farcs8 & quad - \citetalias{PaperVI} \\
GraL J0818$-$2613 & 08:18:28.3 & $-$26:13:25 & 2.164 & 100.1 & 6\farcs2 & quad - \citetalias{PaperVI} \\
GraL J1131$-$4419 & 11:31:00.0 & $-$44:20:00 & 1.090 & 70.4  & 1\farcs6 & quad - \citetalias{PaperIV} \\
GraL J1651$-$0417 & 16:51:05.3 & $-$04:17:25 & 1.451 & 7.3  & 10\farcs1 & quad - \citetalias{PaperVI} \\
GraL J1719+1515   & 17:19:22.6 &   +15:15:46 & 1.716 & \nodata  & 1\farcs1 & double - \citetalias{PaperV} \\
GraL J1817+2729   & 18:17:30.8 &   +27:29:40 & 3.074 & 19.0  & 1\farcs8 & quad - \citetalias{PaperVI} \\
GraL J2017+6204   & 20:17:49.1 &   +62:04:43 & 1.724 & 14.7  & 0\farcs7 & quad - \citetalias{PaperVI} \\
GraL J2103$-$0850 & 21:03:29.0 & $-$08:50:49 & 2.455 & 13.3  & 1\farcs0 & quad - \citetalias{PaperVI} \\
GraL J2200+1448   & 22:00:15.6 &   +14:49:00 & 1.115 & \nodata   & 2\farcs5 & double - \citetalias{PaperV} \\
\hline
SDSS J1141$-$0436 & 11:41:03.9 & $-$04:36:51 & 1.647  &  \nodata &  \nodata & unlensed - \citetalias{PaperVI} \\
\enddata
\tablenotetext{a}{Adopted magnification based on ${\rm SIS}+\gamma$ models presented in \citetalias{PaperVI} (Quads).}
\tablenotetext{b}{For the quads, separation corresponds to the maximum separation.}
\end{deluxetable*}

Strong gravitational lensing, wherein a distant object is magnified and possibly resolved into multiple images by a massive foreground structure, is an extremely valuable tool for studying the universe (see \citealt{2010ARA&A..48...87T} for a review). Not only do strong lenses enable mass reconstruction from the scales of galaxy clusters \citep[e.g.,][]{2018ApJ...863..154P} to the scales of galaxy subhaloes \citep[e.g.,][]{2012Natur.481..341V}, but strong lensing measurements have enabled tests of fundamental physics and cosmology. Using spatially-resolved kinematic measurements of lensed arcs, \citet{2018Sci...360.1342C} tested the predictions of general relativity in the strong-gravity regime. Furthermore, a number of works have exploited time delays between individual images to calculate $H_0$ \citep[e.g.,][]{2017MNRAS.468.2590S, 2018MNRAS.481.1115C, 2018ApJ...853L..31V}.

Of particular importance in the strong lensing regime are background quasars lensed by galaxy-scale masses. Quasar microlensing directly constrains the stellar mass fraction at the position of lensed images, enabling kinematics-independent derivations of dark matter fractions in galaxies \citep{2011ApJ...731...71B, 2014MNRAS.439.2494O}. Furthermore, using flux measurements of lensed quasars to model the mass distribution of lensing galaxies, works including those of \citet{2020MNRAS.492L..12G} and \citet{2020MNRAS.492.5314N} have constrained the characteristics of dark matter structures. And, building on the work of \citet{2004ApJ...610...69K}, which showed that microlensing time delays can enable a measurement of lensed source sizes, multiple works have exploited lensing to measure the properties of quasars \citep[e.g.,][]{2005MNRAS.359..561W, 2007ApJ...661...19P}. Due to the vast utility of these sources, lensed quasars have remained compelling targets for discovery.

While the first lensed quasars were discovered by serendipity \citep{1979Natur.279..381W}, exploiting the full potential of these systems requires both large samples and systematic searches. To that end, the \textit{Gaia} Gravitational Lenses working group (GraL) has exploited the exquisite astrometric precision of the \textit{Gaia} mission \citep{2016A&A...595A...1G} to identify candidate lensed quasars \citep[][Paper I]{PaperI}. \citet[][Paper II]{PaperII} expanded on this work by creating an exhaustive list of known quasars and integrating in the sub-milliarcsecond astrometry of \textit{Gaia} Data Release 2 \citep{2018A&A...616A...1G}. Following refinement of the candidate selection algorithms \citep[][Paper III]{PaperIII} and a demonstration of the potential for \textit{Gaia} observations alone to constrain mass models \citep[][Paper IV]{PaperIV}, \citet[][Paper V]{PaperV} and \citet[][Paper VI]{PaperVI} spectroscopically confirmed a set of doubly and quadruply imaged quasars, respectively. All told, over two dozen lensed quasars have been identified and confirmed by GraL, which is one of several ongoing searches for lensed quasars \citep{2017MNRAS.465.4325O,2018MNRAS.473L.116O,2018MNRAS.479.4345A,2018MNRAS.481.1041T,2018MNRAS.479.5060L,2019MNRAS.483.4242L,2020MNRAS.494.3491L,2019A&A...632A..56K,2020ApJ...899...30L}.

\begin{figure*}
\centering
\vspace{5mm}
   \includegraphics{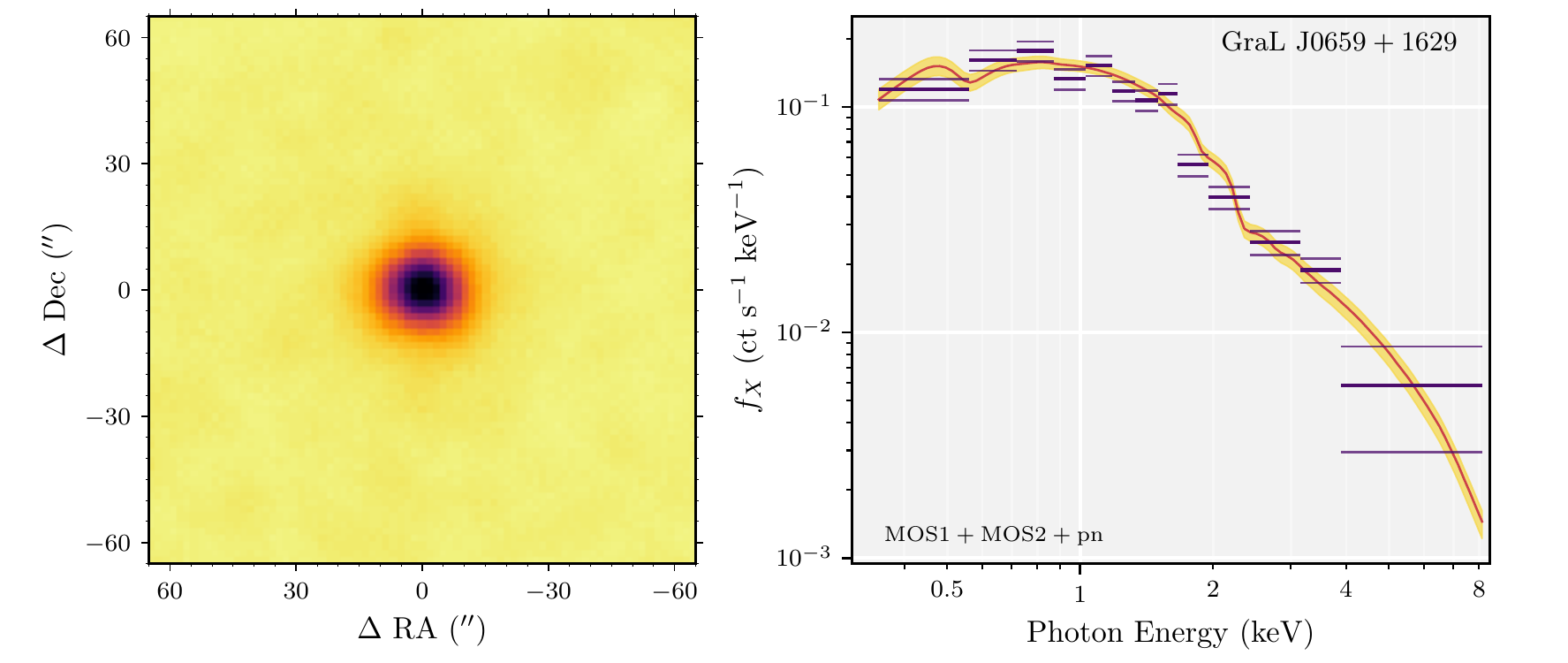}
      \caption{\small EPIC observations of GraL J0659$+$1629. {\bf Left}: imaging observations in the combined three EPIC cameras, covering 0.3--8.0 keV. This image is centered on GraL J0659$+$1629 and has been smoothed with a Gaussian kernel of width $\sigma=4\farcs0$ and binned to pixels of width $1\farcs75$. {\bf Right}: Combined background-subtracted spectra from the three EPIC instruments, binned for plotting purposes, overlaid on the best-fit model and its $1\sigma$ uncertainties (yellow). More detailed versions of this figure for the entire sample are presented in the Appendix in Figures \ref{fig:apx0659} and \ref{fig:apx_upperlimits}.
      }\label{fig:DemoFig}
\end{figure*}

As new gravitational lens systems are discovered, they become intriguing targets for X-ray studies, enabling discoveries beyond that of optical investigations. In particular, as noted by \citet{2012ApJ...744..111P}, a quiescent lensing galaxy does not contribute a meaningful amount of X-ray flux, allowing for improvements in mass modeling. Utilizing the inherent ability to obtain spectral information from each resolved image, \citet{2015ApJ...805..161W}, building on measurements at lower redshifts \citep{2014Natur.507..207R, 2014ApJ...792L..19R}, demonstrated the ability of X-ray observations of lensed quasars to measure black hole spins in the $z>3$ universe. Differences between X-ray and optical light curves have enabled multiple measurements of the size of the X-ray emitting region in lensed quasars \citep[e.g,.][]{2008ApJ...689..755M, 2010ApJ...709..278D, 2013ApJ...769...53M}. In addition, while distant quasars can still be well-studied from optical to radio wavelengths \citep[e.g.,][]{2021ApJ...909...80B}, even luminous quasars with deep X-ray observations are so photon limited as to preclude all but the most basic of spectral analyses \citep[e.g.,][]{2019ApJ...887..171C}; as such, the boost in flux caused by lensing that enables more detailed studies of individual objects is critical in the X-ray regime for exploring the $z>3$ universe.

Of particular interest are microlensing events when lensed objects touch a caustic. These caustic crossing events have been previously observed to produce magnifications in excess of $\times2000$ \citep{2018NatAs...2..334K}. While such extreme magnification events are uncommon and generally associated with the macrocaustics of galaxy clusters \citep{2019A&A...625A..84D}, even smaller-magnification microlensing events could enable studies of distant quasars at a level of detail otherwise only obtainable in the local universe \citep[e.g.,][]{2018MNRAS.475.1925T}. As the strength of a caustic crossing event increases with decreasing source size, the relatively small scale of X-ray emitting regions makes this energy band ideal for exploiting these incidents. \citet{2011ApJ...738...96M} found that the median Einstein radius crossing time for a sample of 87 lensed quasars, which is equivalent to the rate of caustic crossing events, was once per 20 years per lensed image. Due to the rarity of these events, analyses often rely on photometric monitoring of low-magnification events \citep[e.g.,][]{2018A&A...609A..71C, 2018ApJ...869..132F} or of statistical analyses of higher-magnification events \citep[e.g.,][]{2018NatAs...2..324R}. However, with time-domain surveys such as the Zwicky Transient Facility \citep{2019PASP..131a8002B} providing deep coverage of large fractions of the sky at near-daily cadence, we may soon be able to detect caustic-crossing events early and often enough to enable target of opportunity observations. Preliminary X-ray observations are necessary to prioritize these optically-selected events for X-ray follow-up.

In this article we present the X-ray observations and analysis of a subset of the \textit{Gaia} GraL sample. The paper is structured as follows: we present our observations and their reduction in Section \ref{sec:observations}, provide the results in Section \ref{sec:results}, discuss these results in the broader context of ongoing studies in Section \ref{sec:discussion}, and summarize this effort in Section \ref{sec:summary}. Throughout this work, we adopt a flat cosmology with $H_0 = 70\,\textrm{km\,s}^{-1}\,\textrm{Mpc}^{-1}$, $\Omega_M = 0.3$, and $\Omega_\Lambda = 0.7$. All uncertainties are given at the $1\sigma$ level and all upper limits correspond to $3\sigma$ values. Except where otherwise stated, all luminosities presented in this work are not corrected for the lensing magnification.

\section{X-Ray Observations and Analysis}\label{sec:observations}
We proposed a snapshot survey using \xmm\ to observe a sample of 19 lensed quasars from GraL (PropID: 086462, PI: Stern), though this survey was accepted as C Priority, and so only a random sub-sample was observed. In total, ten new quasars were observed in this program, the details of which are given in Table \ref{table:sample}.  One unlensed quasar -- SDSS J1141$-$0436 -- was included in this sample. Though initial  reductions of follow-up spectroscopy suggested a lensed quasar, subsequent analysis revealed this source to be an asterism composed of a Galactic star and a quasar \citepalias{PaperVI}.

Each target was observed for around 10 ks with the European Photon Imaging Camera (EPIC) on \xmm\ \citep{2001A&A...365L...1J}, consisting of two MOS cameras \citep{2001A&A...365L..27T} and a pn CCD camera \citep{2001A&A...365L..18S}. Sources were positioned at the standard EPIC-pn prime position, ensuring they were away from pn chip edges. Camera readout was conducted in full frame mode, and we used the thin optical blocking filter.

\begin{deluxetable*}{clcDrrrc}
\tablecaption{Observations and Fluxes}
\label{table:fluxes}
\tablehead{
\colhead{} &
\colhead{} &
\colhead{} &
\twocolhead{ } &
\multicolumn{3}{c}{Count Rate\tablenotemark{a}} &
\colhead{} \\
\colhead{Target} &
\colhead{OBSID} &
\colhead{Start Date} &
\twocolhead{Exp.} &
\colhead{MOS1} &
\colhead{MOS2} &
\colhead{pn} &
\colhead{$F_{0.3-8.0}$} \\
\colhead{} &
\colhead{} &
\colhead{(YYYY-mm-dd)} &
\twocolhead{(ks)} &
\multicolumn{3}{c}{(${\rm ct}\ {\rm ks}^{-1}$)} &
\colhead{($10^{-14}\ {\rm erg}\ {\rm cm}^{-2}\ {\rm s}^{-1}$})}
\decimals
\startdata
GraL J0659+1629   & 0864620401                    & 2021-Apr-07 & 16.7 & $55.5 \pm  2.7 $& $44.8 \pm  2.4 $& $173 \pm 11 $& $ 56.3^{+  3.2}_{-  3.0} $\\
GraL J0818$-$2613 & 0864620501                    & 2020-Oct-23 &  8.5 & $26.5 \pm  2.0 $& $21.7 \pm  1.9 $& $60.8 \pm  4.5 $& $ 38.0^{+  4.3}_{-  4.0} $\\
GraL J1131$-$4419 & 0864620701\tablenotemark{b,c} & 2020-Dec-11 & 16.1 & $33.1 \pm  3.7 $& $32.1 \pm  3.3 $& \nodata & $ 33.2^{+  4.7}_{-  3.5} $\\
GraL J1651$-$0417 & 0864621301\tablenotemark{b,c} & 2021-Mar-17 & 17.4 & $36.0 \pm  2.1 $& $39.4 \pm  2.1 $& \nodata & $ 42.1^{+  2.6}_{-  2.7} $\\
GraL J1719+1515   & 0864622501\tablenotemark{c}   & 2021-Mar-02 & 11.0 & $ 8.9 \pm  1.5 $& $13.4 \pm  1.7 $& $39.4 \pm  4.5 $& $ 11.8^{+  1.8}_{-  1.4} $\\
GraL J1817+2729   & 0864621501\tablenotemark{b,c} & 2020-Oct-24 & 19.5 & $< 6.1$ & $< 4.4$ & \nodata & $<17.30$\\
GraL J2017+6204   & 0864621701                    & 2020-Jul-09 & 15.3 & $ 6.9 \pm  0.9 $& $ 5.7 \pm  0.8 $& $17.6 \pm  2.3 $& $ 11.8^{+  2.1}_{-  1.8} $\\
GraL J2103$-$0850 & 0864621901\tablenotemark{c}   & 2020-Nov-05 & 14.2 & $11.5 \pm  1.4 $& $12.6 \pm  1.3 $& \nodata & $ 15.5^{+  2.0}_{-  2.1} $\\
GraL J2200+1448   & 0864622001\tablenotemark{c}   & 2020-Nov-17 & 12.3 & $ 6.7 \pm  1.5 $& $ 8.4 \pm  1.4 $& \nodata & $8.59^{+1.74}_{-1.45} $\\
\hline
SDSS J1141$-$0436 & 0864620801\tablenotemark{b,c} & 2020-Dec-25 & 15.7 & $<10.9$ & $< 8.2$ & \nodata & $<16.52$\\
\enddata
\tablenotetext{a}{Background-subtracted count rate from 0.3--8.0 keV.}
\tablenotetext{b}{Affected by radiation.}
\tablenotemark{c}{pn experienced full scientific buffer during observation.}
\end{deluxetable*}

Observations were conducted from 2020 July to 2021 April; details of these observations are given in Table \ref{table:fluxes}. We note that there are three additional OBSIDs associated with our program that are not included in this analysis. Two of these observations (0864622301 and 0864622401) were conducted with the EPIC filter wheel in the closed position due to enhanced radiation at the start of a revolution, while the third (0864621401) was affected by radiation at such a significant amount as to be unusable. Several other observations were also affected by radiation, as indicated in Table \ref{table:fluxes}, but at a level that still allowed the data to be usable. We also note those sources for which the pn camera experienced a full scientific buffer. Normally caused by a high radiation background, the full scientific buffer causes the pn camera to switch to counting mode, thereby no longer recording scientific data. 

We reduced and processed these observations using the Scientific Analysis System (SAS, \citealt{2004ASPC..314..759G}) v19.0.0. To standardize our analysis, we used the \texttt{xmmextractor} script to produce event files and extract spectra. As part of this analysis, we adopted standard analysis flags (\texttt{PATTERN}${\leq}12$ for MOS and \texttt{PATTERN}${\leq}4$ for pn) and good time intervals (\texttt{RATE}${\leq}0.35$ for MOS and \texttt{RATE}${\leq}0.4$ for pn). Source spectra were extracted in \texttt{xmmextractor}-selected apertures, while background spectra were extracted from off-source circular apertures of varying size.

Spectral fitting was performed using the python implementation of \texttt{XSPEC} v12.11.1 \citep{1996ASPC..101...17A}. We used a simple absorbed power-law model to fit our sources (\texttt{phabs}$\times$\texttt{powerlaw}). For all targets, we adopt a Galactic neutral Hydrogen column density, $N_{\rm H}$, based on the \ion{H}{1} HI4PI Survey \citep{2016A&A...594A.116H}. Both the normalization of the power law and the photon index, $\Gamma$, were free to vary. We fixed the spectra of all three EPIC cameras to the same normalization, as studies with significantly deeper spectra have found that cross-normalization terms are effectively unity \citep[e.g.,][]{2014A&A...564A..75R, 2015MNRAS.453.3953L}. We binned our spectra to a minimum of only one count per bin, and we therefore used the modified $C$-statistic to evaluate best-fits and errors \citep{1979ApJ...228..939C, 1979ApJ...230..274W}. 

Additionally, we fit each source including an absorption component at the quasar redshift (\texttt{phabs}$\times$\texttt{powerlaw}$\times$\texttt{zphabs}). Two sources have a redshifted column density, $N_{H,z}$, that is not consistent with 0; for the other objects, we only report the results of the simpler fits. These two sources -- GraL J0818$-$2613 and GraL J2017$+$6204 -- were previously identified in \citetalias{PaperVI} as having optical spectral signatures of absorption. GraL J0818$-$2613 has a red continuum and weak Ly$\alpha$ emission, while GraL J2017$+$6204's spectrum is reddened with broad absorption line (BAL) features. The only other quasar in our sample with optical features of absorption is GraL J1817$+$2729, which is not strongly detected in our observations.

\begin{deluxetable*}{cccDcccc}
\tablecaption{Mid-IR luminosities and X-ray properties of the sample.}
\label{table:xray_fits}
\tablehead{
\colhead{Target} &
\colhead{$\log \nu L_{\rm 6 \mu m}$} &
\colhead{$N_{\rm H}$} &
\twocolhead{norm\tablenotemark{a}} &
\colhead{$\Gamma$} &
\colhead{$N_{\rm H,z}$} &
\colhead{$\log L_{2-10}$} &
\colhead{$C$/DOF}\\
\colhead{} &
\colhead{(${\rm erg}\ {\rm s}^{-1}$)} &
\colhead{($10^{20}\, {\rm cm}^{-2}$)} &
\twocolhead{($10^{-5}$)} &
\colhead{} &
\colhead{($10^{22}\, {\rm cm}^{-2}$)} &
\colhead{(${\rm erg}\ {\rm s}^{-1}$)} &
\colhead{} }
\decimals
\startdata
GraL J0659$+$1629 & $46.81^{+0.03}_{-0.04}$ & 11.60 & $12.83^{+0.63}_{-0.65}$ & $1.87^{+0.07}_{-0.07}$ & \nodata & $46.44^{+0.02}_{-0.02}$ &  831.18/937 \\
GraL J0818$-$2613 & $47.54 \pm 0.01$ & 13.40 & $6.76^{+2.92}_{-1.92}$ & $1.42^{+0.26}_{-0.25}$ & $8.07^{+4.21}_{-3.64}$ & $45.89^{+0.11}_{-0.10}$ &  520.04/623 \\
GraL J1131$-$4419 & $45.97 \pm 0.02$ & 4.86 & $7.14^{+0.79}_{-0.83}$ & $1.96^{+0.20}_{-0.19}$ & \nodata & $45.09^{+0.06}_{-0.06}$ &  282.68/322 \\
GraL J1651$-$0417 & $45.74^{+0.08}_{-0.10}$ & 9.52 & $9.31^{+0.53}_{-0.63}$ & $1.88^{+0.09}_{-0.09}$ & \nodata & $45.53^{+0.02}_{-0.03}$ &  568.31/682 \\
GraL J1719$+$1515 & $46.20 \pm 0.02$ & 5.44 & $2.63^{+0.28}_{-0.30}$ & $1.99^{+0.17}_{-0.17}$ & \nodata & $45.13^{+0.05}_{-0.05}$ &  290.77/278 \\
GraL J1817$+$2729 & $47.07 \pm 0.02$ & 8.43 & $<1.59$ & \nodata & \nodata & $<45.56$ & \nodata  \\
GraL J2017$+$6204 & $46.22 \pm 0.02$ & 13.40 & $2.53^{+2.81}_{-1.19}$ & $1.49^{+0.52}_{-0.44}$ & $10.96^{+9.00}_{-6.50}$ & $45.24^{+0.21}_{-0.17}$ &  371.22/392 \\
GraL J2103$-$0850 & $46.76^{+0.03}_{-0.04}$ & 6.02 & $2.75^{+0.42}_{-0.31}$ & $1.68^{+0.17}_{-0.19}$ & \nodata & $45.57^{+0.05}_{-0.05}$ &  292.51/317 \\
GraL J2200$+$1448 & $45.17^{+0.09}_{-0.11}$ & 4.31 & $2.25^{+0.33}_{-0.39}$ & $2.43^{+0.36}_{-0.35}$ & \nodata & $44.46^{+0.12}_{-0.13}$ &  267.07/279 \\
\hline
SDSS J1141$-$0436 & $45.54^{+0.12}_{-0.16}$ & 3.05 & $<2.38$ & \nodata & \nodata & $<45.04$ & \nodata   \\
\enddata
\tablenotetext{a}{Normalization of the \texttt{powerlaw} component, with units ${\rm photons}\ {\rm s}^{-1}\ {\rm cm}^{-2}\ {\rm keV}^{-1}$ at 1 keV.}
\end{deluxetable*}

We computed the uncertainties on fit parameters by measuring contours in the $C$ statistic. As noted by \citet{1979ApJ...228..939C}, $\Delta C$ behaves as $\Delta \chi^2$ when evaluating confidence intervals, so that the $1\sigma$ uncertainties include those fits where $\Delta C \leq 2.30$ \citep[or $\Delta C \leq 3.53$ for the three-component model, e.g.,][]{1976ApJ...208..177L}. Figure \ref{fig:DemoFig}, presenting GraL~J0659$+$1629, shows an example of our reduced data. The combined image from the three EPIC cameras are shown in the left, smoothed with a Gaussian kernel of width $\sigma=4\farcs0$ and with individual normalizations adjusted for presentation purposes. In the right, we show the background-subtracted combined count rate spectra from all three cameras, binned for plotting purposes. Horizontal bars show the source count rate, with thinner bars above and below corresponding to the $1\sigma$ uncertainties, using the methods of \citet{1986ApJ...303..336G}. The best fit model and its $1\sigma$ uncertainties were folded through the spectral responses with \texttt{XSPEC} and are plotted with the red line and yellow region, respectively. More detailed versions of this figure are presented for each quasar in our sample in the Appendix as Figures \ref{fig:apx0659} and \ref{fig:apx_upperlimits}.

\section{Results}\label{sec:results}
In this section we present the results of our analysis of the ten observed quasars. We first present the observed properties for the entire sample in \S~\ref{ssec:full_results}, including notes on specific parameters. Then, in \S~\ref{ssec:individuals}, we discuss individual quasars, including how results may be influenced by the observing conditions.

\subsection{Full Sample} \label{ssec:full_results}
We begin our analysis of these quasars with measurements of their flux. As with all results reported here, we do not attempt to differentiate the properties of individual point quasar images, as the \xmm\ EPIC half energy width is ${\sim}15^{\prime\prime}$. One motivation for reporting flux values is to facilitate the planning of future observations of these sources in the event of a caustic crossing event. As these observations may be conducted with either \textit{Chandra} or \xmm, we present flux values in the range of 0.3--8.0 keV, which is a suitable broad baseline for both observatories. Total energy fluxes, in units of ${\rm erg}\ {\rm cm}^{-2}\ {\rm s}^{-1}$, are given in Table \ref{table:fluxes} for all sources. These flux values are derived in \texttt{XSPEC} from the best-fitting model fits. For those quasars that were not well-fit, $3\sigma$ upper limits are given instead. We also report the background-subtracted source count rates in Table \ref{table:fluxes}. Upper limits are again given for rates not detected at a $3\sigma$ level, and we do not report values for sources observed for less than 1.5 ks in a camera.

Next, we present the fitted X-ray properties of these quasars. The normalization and photon index of each object's power law component is listed in Table \ref{table:xray_fits}. As with flux measurements, we present the upper limits on the normalization for the two quasars that were not detected. We also present the unobscured, rest-frame 2--10 keV luminosities. These model-derived luminosities assume the source is entirely composed of power-law emission in this energy band. For the two quasars that were well-fit by this model, we also include the constraints on the redshifted column density.

\begin{figure*}
\centering
\vspace{5mm}
   \includegraphics{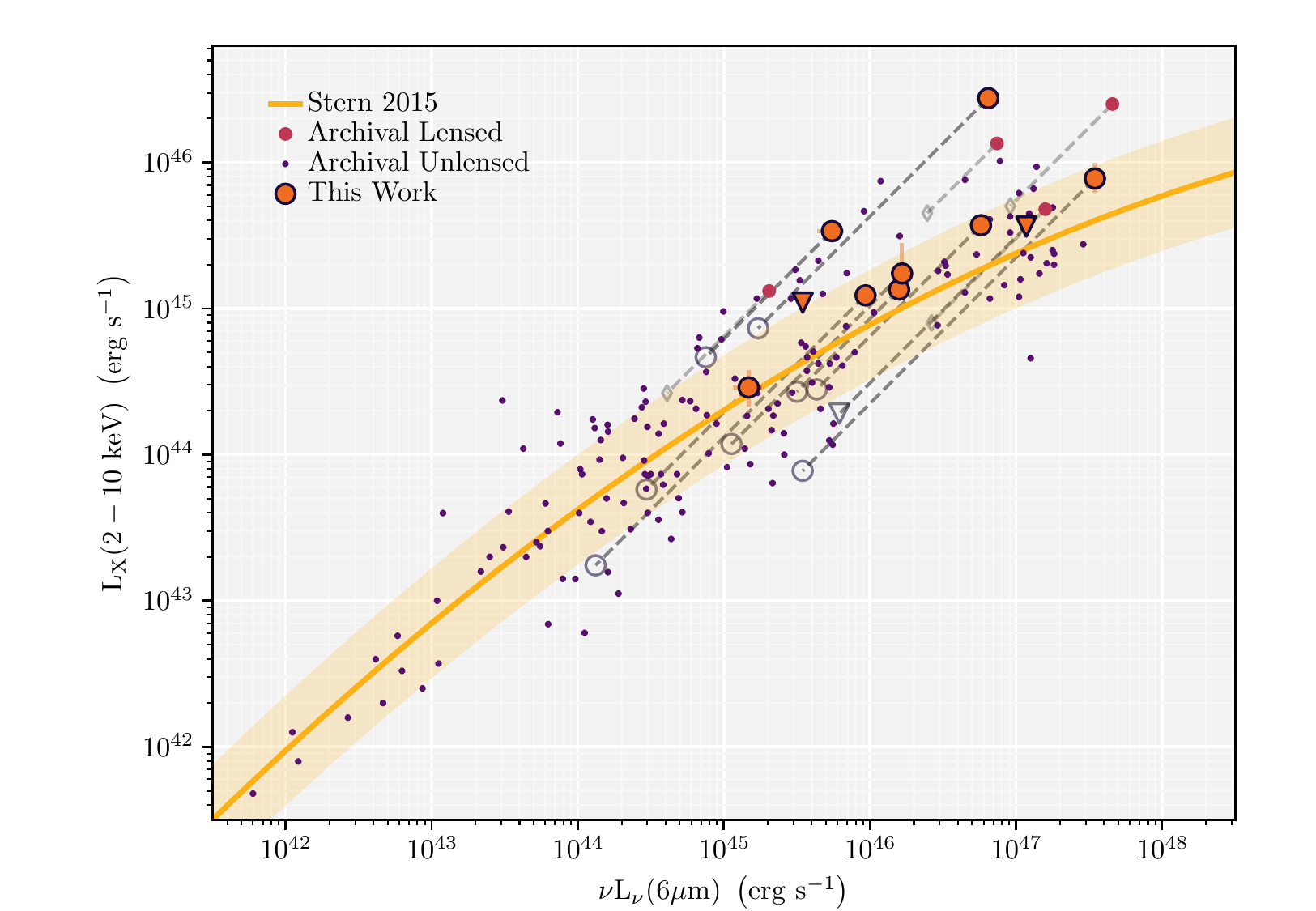}
      \caption{Rest-frame, absorption-corrected 2–10~keV X-ray luminosity against rest-frame $6 \mu$m luminosity for the quasars analyzed in this work (orange), as well as a sample of archival lensed and unlensed AGN. Upper limits are indicated by downward pointing triangles. The X-ray to mid-infrared luminosity relation of \citet{Stern:15} is shown by the yellow line. For all lensed quasars, the dashed gray lines indicate their unmagnified luminosities; we assume a magnification of $\mu=5$ for all quasars without reported magnifications. The sources of archival values are described in the text.
      }\label{fig:lx-lir}
\end{figure*}

Finally, we include the rest-frame 6 $\mu$m luminosity for each of these objects. Following, e.g., \citet{Stern:15}, we use photometry from the \textit{Wide-field Infrared Survey Explorer} \citep[\textit{WISE};][]{2010AJ....140.1868W} and known redshifts from \citetalias{PaperIV}, \citetalias{PaperV}, and \citetalias{PaperVI} to calculate $\nu L_{6 \mu {\rm m}}$. For our entire sample, rest-frame 6 $\mu$m lies between the \textit{WISE} W3 (12 $\mu$m) and W4 (24 $\mu$m) channels, and we compute luminosities through linear interpolation of these values. IR luminosities are listed in Table \ref{table:xray_fits}. Unlike the X-ray measurements, which are expected to only have minimal contamination from the lensing galaxy, these values could potentially be slightly boosted in flux due to the contribution of the intervening galaxy. On the other hand, lensing preferentially occurs from more massive, i.e., early-type, galaxies, which have falling spectral energy distributions beyond rest-frame $H$-band, so the expectation is that the W3 and W4 flux from these systems is dominated by the lensed quasar emission.

We show the distribution of X-ray to IR luminosities for this sample in Figure \ref{fig:lx-lir}. For all lensed quasars, we also plot a magnification track, showing what these values would be were the quasar unlensed. For the quad lenses, we use the modeled magnification values from \citetalias{PaperVI} (listed in Table \ref{table:sample}), while we adopt a value of $\mu=5$ for the doubly-imaged lenses \citep[a typical value for these systems, e.g.,][]{2000ApJ...535..692K, 2016MNRAS.458....2R}. Also shown is the relation between X-ray and IR luminosities presented by \cite{Stern:15}. While linear at lower luminosities, this relation has a characteristic flattening above $\nu L_{6 \mu {\rm m}}{\sim}10^{44}\ {\rm erg}\ {\rm s}^{-1}$, believed to be caused by the X-ray emission saturating as the corona cools and softens with increasing thermal emission from the disk \citep[e.g.,][]{2013MNRAS.433.2485B}. We also include a sample of archival lensed quasars (\citealt{Just:07}, \citealt{Stern:20}, and Walton et al., submitted), local Seyferts \citep{Horst:08, Gandhi:09}, and luminous quasars \citep{Just:07}. For the lensed quasar sample, magnification tracks are for reported values if known, and are otherwise also assumed to be $\mu=5$.

\subsection{Notes on Individual Quasars} \label{ssec:individuals}

\subsubsection{GraL J0659+1629}
The highest redshift quasar in our sample, GraL J0659+1629 is also the X-ray brightest. Consequently, this quasar is the most X-ray luminous object in our sample by almost an order of magnitude. \citetalias{PaperVI} reported that there are no archival radio sources associated with this source, and the closest object in the 3 GHz Very Large Array Sky Survey Epoch 1 Quick Look catalog is almost two arcminutes away \citep{2020RNAAS...4..175G}. We have also carried out deeper VLA observations (D. Dobie et al., in prep.) and detected radio sources at the location of all four optical images with a typical flux density of ${\sim}90\ \mu{\rm Jy}$, comparable to the optical flux density reported by \citetalias{PaperVI}. This quasar therefore does not fit the standard definition of radio-loud, i.e. having a radio flux density greater than ten times its optical flux density \citep[e.g.,][]{1989AJ.....98.1195K}. Conversely, the most X-ray luminous quasars in the $z>3$ universe tend to be radio-loud quasars or blazars \citep[e.g.,][]{2021AstL...47..123K}. As such, this source presents a unique opportunity to study the radio-quiet $z>3$ quasar population in detail. We also note that the high observed X-ray luminosity is potentially indicative of a large magnification; as discussed by \citet{Stern:15}, X-ray luminosities tend to saturate above ${\sim}10^{44}\ {\rm erg}\ {\rm s}^{-1}$, so the observed value of $10^{46.4}\ {\rm erg}\ {\rm s}^{-1}$ is the result of either a very intrinsically luminous quasar or a large luminosity boost from lensing. From the modeling presented in \citetalias{PaperVI}, $\mu=37.6$, demonstrating the validity of this technique for identifying significantly lensed quasars.

\begin{figure}
\centering
\vspace{5mm}
   \includegraphics{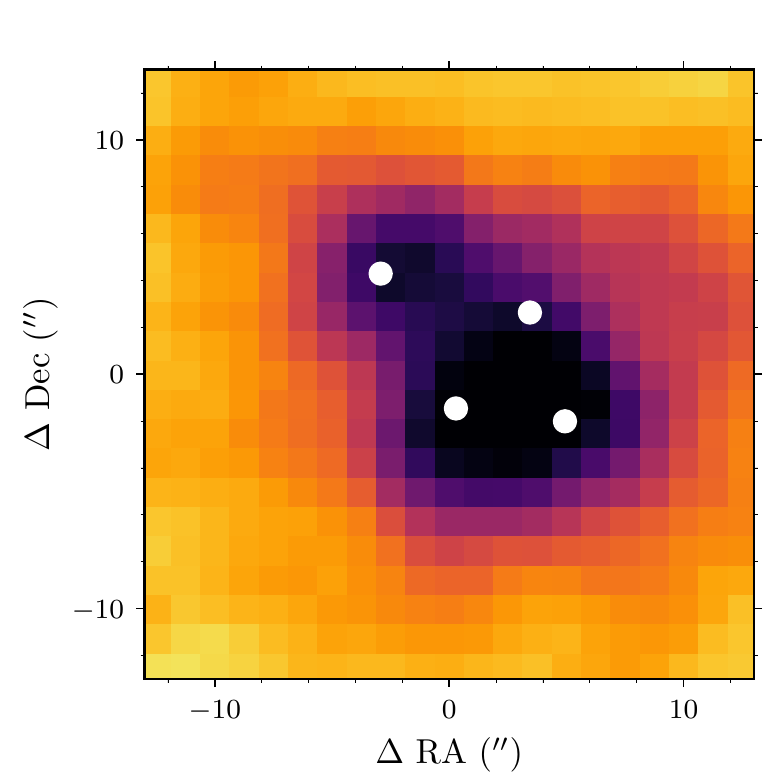}
      \caption{MOS1+MOS2 0.3--8.0 keV image of GraL J1651$-$0417, smoothed with a Gaussian kernel of width $1\farcs5$. The position of the lensed sources, as identified by \citetalias{PaperVI}, are indicated by white dots. A small relative offset has been applied to the lensed image positions in this figure, in keeping with the expected pointing accuracy of \xmm. Although the extended wings of these sources overlap, the most distant lensed source can nevertheless be resolved.
      }\label{fig:resolved_source}
\end{figure}

\subsubsection{GraL J1131--4419}

\citet{2000IAUC.7432....3V} report an X-ray source at this position in \textit{ROSAT} All-Sky Survey observations, 1RXS J113058.9$-$441949. While that catalog only reports a source count rate, the second \textit{ROSAT} all-sky source catalog \citep{2016A&A...588A.103B} includes properties from a power-law spectral fit. The reported absorption-corrected flux in the 0.1-2.4 keV band is $1.6 \times 10^{-11}\ {\rm erg}\ {\rm s}^{-1}\ {\rm cm}^{-2}$, which is almost two orders of magnitude brighter than what we report here. Some of this discrepancy can be explained by \citet{2016A&A...588A.103B} effectively adding in flux by correcting for absorption and the differences in energy bands. Further differences may be driven by the fitted power-law, which has a best-fit photon index of $\Gamma=3.07$ for the \textit{ROSAT} data. As discussed by \citet{2020ApJ...900..189C}, an excessively steep fit to the photon index caused by limited source counts can effect a larger calculated flux at soft energies. Yet $\Gamma$ cannot be entirely to blame, as the normalization, $n_{\rm 2RXS}=(9.4\pm5.4) \times 10^{-4}$, is still an order of magnitude larger than what we find in the more recent \xmm\ observations. 

We also note that another X-ray source is seen in the new X-ray imaging, roughly $50^{\prime\prime}$ to the north at the location of the high proper motion star 2MASS J11310001$-$4419088. However, it is unlikely that this is the source of the large \textit{ROSAT} flux for two reasons. First, although the separation of the two objects may lead to some flux contamination in the \textit{ROSAT} imaging \citep[see][]{2000A&AS..141..507B}, the star is fainter than GraL J1131$-$4419 in the new observations and much fainter than the earlier flux value. Second, 1RXS J113058.9$-$441949 is only $16^{\prime\prime}$ from GraL J1131$-$4419, consistent with the expected positional uncertainty \citet{2016A&A...588A.103B} report for \textit{ROSAT} coordinates. In contrast, the \textit{ROSAT} detection is $42^{\prime\prime}$ from 2MASS J11310001$-$4419088, implying that the star was not the source of the X-ray flux. There are no further bright X-ray objects within $5^\prime$ of the lensing system.

It is not clear what is responsible for such a change in the observed flux. While AGN are known to have intrinsic flux variations in X-rays \citep[e.g.,][]{2004ApJ...611...93P}, the observed dimming is too large to be explained by stochastic variability from changes in black hole fueling alone \citep{2018MNRAS.476L..34S}. Such a large dimming over 30 years (15 years in the source frame) could be attributed to the quasar being a changing-look AGN \citep{2020ApJ...898L...1R}, although this is difficult to assess without a spectrum from the earlier epoch. Serendipitous \textit{Swift} observations from 2009-2012 show no significant difference in the X-ray flux with what is found here ($f_{0.3-10.0\ {\rm keV}} = 6^{+1}_{-2}\times 10^{-13}\ {\rm erg}\ {\rm s}^{-1}\ {\rm cm}^{-2}$; \citealt{2020ApJS..247...54E}). Alternatively, the \textit{ROSAT} observations could have coincided with a microlensing event, although this, too, would be an extreme value for such an effect \citep{2012ApJ...755...24C}.

\subsubsection{GraL J1651--0417}

This quasar has the largest maximum separation of our sample, at $10\farcs1$ \citepalias{PaperVI}. The most separated lensed image is located to the NE, while the three other images in this quad are located in close proximity to each other. In the X-ray observations of this system, presented in Figure \ref{fig:resolved_source}, we find that the quasar is composed of two separate sources, with the second source appearing in the direction and at the separation expected of the NE image. Individual lens images have been resolved by \xmm\ when the lensing object is a galaxy cluster \citep[e.g.,][]{2006A&A...454..493L}, but previous observations of sources with galaxy-scale lenses have heretofore been unresolved with this observatory \citep[e.g.,][]{2008A&A...490..989F, 2016ApJ...824...53C}. GraL J1651$-$0417 is thus a potentially interesting source for future X-ray studies with large effective area but worse-than-arcsecond resolution, such as X-IFU on \textit{Athena} \citep{2018SPIE10699E..1GB}. 

\subsubsection{GraL J1817+2729}
Despite having one of the highest inferred 6 $\mu$m luminosities of our sample, this gravitational lens system, known as Hercules' Sword \citepalias{PaperVI}, is undetected in a nominal exposure of 19.5 ks. However, this observation was heavily affected by radiation; the pn camera experienced a full scientific buffer and was rendered unusable for our analysis, while the good time intervals for the MOS cameras only summed to 3.1 and 3.0 ks for MOS1 and MOS2, respectively. Nevertheless, the strict upper limits on measured count rates place this $z=3.07$ lensed quasar as the faintest target in our sample. From a mass model of the system, \citet{2019MNRAS.483.4242L} report a magnification for Hercules' Sword of $\mu=14.2^{+1.9}_{-0.9}$, similar to the value of $\mu=19.0$ derived from \citetalias{PaperVI}. Based on that, the unmagnified X-ray luminosity is, at most, of order $10^{44}\ {\rm erg}\ {\rm s}^{-1}$, while the IR luminosity is still approximately $10^{46}\ {\rm erg}\ {\rm s}^{-1}$. This value, even at the X-ray limit, is still a large offset from the \cite{Stern:15} relation.

One potential explanation is that the IR luminosity is contaminated in a way the X-ray measurement is not. \citetalias{PaperVI} spectroscopically identified a Galactic mid-type star ${\sim}2^{\prime\prime}$ NW of the lens, which is the brightest $I$-band object in the system \citep{2018RNAAS...2..187R}. Subaru imaging and associated mass modeling presented by \citet{2018RNAAS...2..187R} show that the lensing galaxy is brighter ($I$-band) than two of the lensed images and is suggestive of an edge-on, dusty disk. Neither a typical Galactic star or an inactive galaxy should be able to mimic such a large IR luminosity, however.

Conversely, the spectrum of this lensed quasar shows strong \ion{C}{4} $\lambda1549$ BAL features. Previous studies have found that the strength of BAL features correlates with a reduced X-ray luminosity \citep{2009ApJ...692..758G}. The presence of this correlation in observed hard-energy \textit{NuSTAR} observations suggests that this faintness is intrinsic, not caused by absorption, and so would still be present even at $z=3$ \citep{2014ApJ...794...70L}. BAL quasars can be more luminous than the limit set for Hercules' Sword -- \citet{2018MNRAS.479.5335V} reported on \xmm\ observations of five $z{\sim}2$, $M_{\rm BH}\sim10^{10}\ M_\odot$ quasars with BAL features, finding luminosities of $L_{2-10}{\gtrsim}10^{45}\ {\rm erg}\ {\rm s}^{-1}$, while \citet{2020ApJ...900..189C} reported on an unlensed $M_{\rm BH} = 3 \times 10^{9}\ M_\odot$, $z=6.59$ BAL quasar with $L_{2-10} \sim 6 \times 10^{44}\ {\rm erg}\ {\rm s}^{-1}$. However, the faintness of GraL J1817+2729 is still in keeping with the expectation of an X-ray weak quasar.

\subsubsection{GraL J2103-0850}
This gravitational lens system is associated with a source detected in the \textit{ROSAT} All-Sky Faint Source Catalog \citep{2000IAUC.7432....3V}, 1RXS J210328.9$-$085039. In the second \textit{ROSAT} all-sky source catalog, \citet{2016A&A...588A.103B} report an absorption-corrected 0.1-2.4 keV flux from an assumed power-law model of $F_{\rm 2RXS}=33 \times 10^{-14}\ {\rm erg}\ {\rm s}^{-1}\ {\rm cm}^{-2}$. Considering the slightly softer energy range of this observation and the correction for absorption, this value is consistent with what we report here, suggesting only a minimal amount of variation since the \textit{ROSAT} observations of 1990/1991. 

\section{Discussion}\label{sec:discussion}

In their analysis of MG 1131+0456, \citet{Stern:20} proposed that the $L_X-\nu L_\nu(6\ \mu{\rm m})$ relation could act as a means to identify lensed quasars. As shown in Figure \ref{fig:lx-lir}, magnification pushes sources on the \citet{Stern:15} relation up and off; thus, any sources with anomalously high X-ray luminosities for their mid-infrared luminosity could indicate lensing. However, most of the sources analyzed here are consistent with the \citet{Stern:15} relation, within the expected scatter. The lack of excess X-ray luminosity is most likely a result of some combination of small magnification factors, intrinsic X-ray luminosities lying below the relation, and the roughly linear correlation at lower luminosities minimizing the impact of magnification on producing deviations. While X-ray luminosity offsets should nevertheless serve as a means of identifying lensing among the most luminous quasars and the strongest magnification lenses, as is demonstrated here by GraL J0659$+$1629, we should not expect the overall population of lensed sources to only be outliers.

One of the motivations for this work was to establish a baseline set of flux measurements of lensed quasars to facilitate future observations of caustic crossing events. The full all-sky survey of \textit{SRG}/eROSITA (eRASS) is expected to reach a point source sensitivity of $f_{eRASS} \lesssim 10^{-14}\ {\rm erg}\ {\rm s}^{-1}\ {\rm cm}^{-2}$ \citep{2021A&A...647A...1P}. As such, we expect all of the lens systems presented here to be detected by the full survey and the brightest of these to potentially have multiple observations to constrain their variability. However, that is the limiting sensitivity for detection; even the simple spectral analyses reported here will be beyond the capabilities of the eRASS. As such, future observations of lensed quasars with \xmm\ and \textit{Chandra} are still warranted.

Another potential advantage of the \textit{SRG}-based observatories comes in their potential to detect X-ray variability. The medium-energy ART-XC telescope on \textit{SRG} is performing daily scans of the sky at 4--12 keV. These scans have sensitivities of ${\sim}2\times10^{-11}\ {\rm erg}\ {\rm s}^{-1}\ {\rm cm}^{-2}$ and survey roughly 1\% of the sky every day \citep{2021arXiv210413267S}. It is possible -- albeit unlikely -- that a caustic crossing event could produce the magnifications necessary to boost one of the lensed quasars in this sample into that flux threshold. As previous transient sources detected by ART-XC have also been seen by eROSITA \citep[e.g.,][]{2020ATel14206....1M,2020ATel14219....1S}, we would expect similar results from the softer survey. As further gravitational lenses are spectroscopically confirmed, archival observations may reveal past extreme magnification events.

Finally, we note the potential for \xmm\ in the study of distant lensed quasars. \textit{Chandra}, with its exquisite angular resolution enabling the separation of individual sources, is often used for studies of lensed quasars \citep[e.g.,][]{2012ApJ...755...24C, 2017ApJ...836..206G, 2020ApJ...894..153D}. However, for faint sources, \textit{Chandra} will be unable to detect the necessary amount of photons for a temporal analysis without deep observations; meanwhile, as demonstrated by \citetalias{PaperII}, \textit{Gaia} observations can provide precise astrometry, obviating that requirement from X-ray observations. In cases such as these, when only spectral information is desired at X-ray energies, \xmm\ is more than suited for the task.

\section{Summary}\label{sec:summary}

We have presented \xmm\ X-ray observations of nine lensed quasars and one unlensed source selected by \textit{Gaia} GraL. Observations were relatively short ($<20$ ks), and represent an exploratory program into the nature of the GraL sample. The primary results of this work are as follows.
\begin{itemize}
    \item We report X-ray fluxes and \xmm\ EPIC count rates for eight of the lensed quasars, as well as upper limits for the ninth. Most sources have fluxes of $F_{0.3-8.0}\approx10^{-13}\ {\rm erg}\ {\rm s}\ {\rm cm}^{-2}$ and count rates of at least 10 ct ${\rm ks}^{-1}$ in each MOS camera. These measurements will be invaluable in planning future targeted observations of caustic crossing events.
    \item Using \texttt{XSPEC}, we fit the observed quasars with an absorbed power-law, and we report the best-fit values of this in Table \ref{table:xray_fits}. From these fits, we also derive rest-frame 2--10 keV unabsorbed luminosities. Here, we find that the observed sample covers over two orders of magnitude in X-ray luminosity.
    \item Despite observing it for almost 20 ks, we do not detect GraL J1817+2729, one of the two most IR luminous quasars in our sample. This is partially due to severe radiation effects during the observation, which cut the effective exposure time to 3 ks in the MOS cameras and which overwhelmed the pn camera entirely. However, the upper limit we infer from these limited observations nevertheless reveals that this quasar is X-ray faint, perhaps related to its observed BAL features. Due to the lensing magnification, deeper observations may enable the first detailed look at an X-ray faint quasar in the early Universe ($z>3$).
    \item We observe GraL J1131$-$4419, which was previously detected in the \textit{ROSAT} All-Sky Survey. The X-ray flux reported from that survey is almost two orders of magnitude brighter than what we find here. As the \textit{ROSAT} observations were taken 30 years prior to the \xmm\ observations, it is not entirely clear what the cause of the variability is, but this could potentially be indicative of a major microlensing event in the older observations.
    \item MOS observations of GraL J1651$-$0417 reveal an extended structure to the NE of the main component of the quasar emission. Using a small smoothing scale, we are able to observe two distinct structures in this lens system, in the orientation expected from the \textit{Gaia}-observed positions of the lensed images. With a maximum separation of $10\farcs1$, this is the most closely-separated gravitational lens system resolved into multiple components by \xmm.
\end{itemize}

{\small The work of TC and DS was carried out at the Jet Propulsion Laboratory, California Institute of Technology, under a contract with NASA. TC's research was supported by an appointment to the NASA Postdoctoral Program at the Jet Propulsion Laboratory, California Institute of Technology, administered by Universities Space Research Association under contract with NASA. DJW acknowledges support from the Science and Technology Facilities Council (STFC) in the form of an Ernest Rutherford Fellowship (grant ST/N004027/1). LD acknowledges support from the ESA PRODEX Programme `{\it Gaia}-DPAC QSOs' and from the Belgian Federal Science Policy Office. SAK was partially supported by the German Aerospace Agency (grant 50QG1402). DS acknowledges support from the European Research Council (ERC) under the European Union’s Horizon 2020 research and innovation programme (grant agreement No 787886).

Based on observations obtained with {\it XMM-Newton}, an ESA science mission with instruments and contributions directly funded by ESA Member States and NASA.}

\vspace{5mm}
\facility{XMM}
\software{PyFITS \citep{1999ASPC..172..483B},
          SAS \citep{2004ASPC..314..759G},
          XSPEC \citep{1996ASPC..101...17A}}

\textcopyright\ 2021. All rights reserved.
\bibliography{bibliography}

\appendix
\section{EPIC Observations and Spectral Fits of the Sample}

\begin{figure*}
\centering
\vspace{5mm}
   \includegraphics{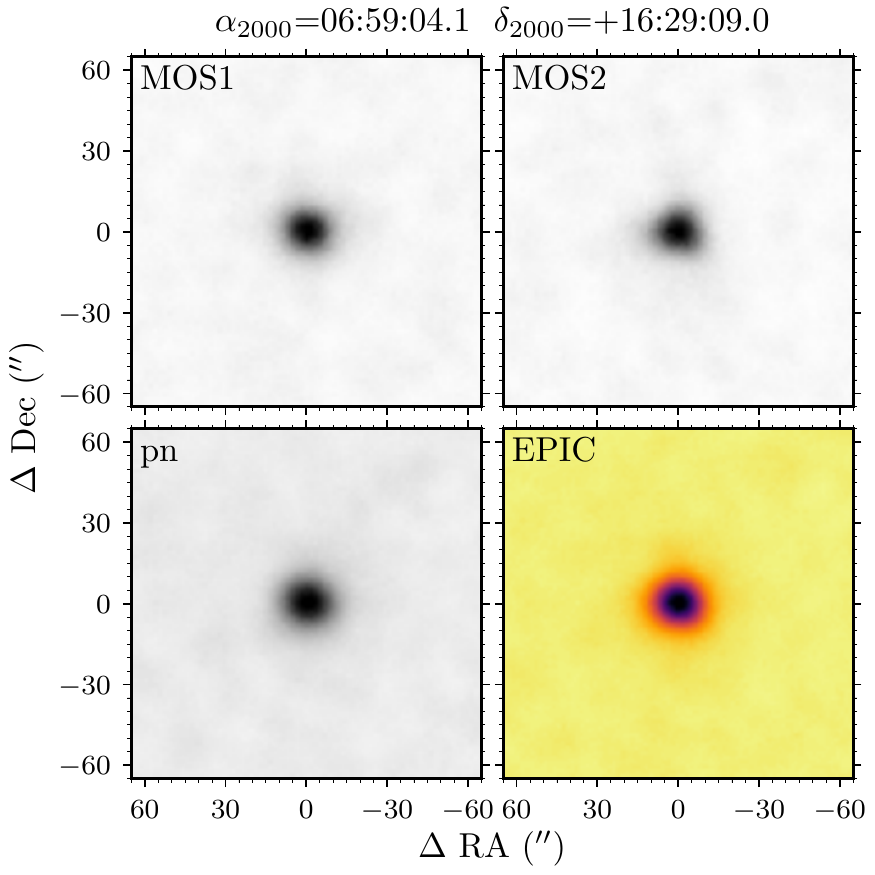}
   \includegraphics{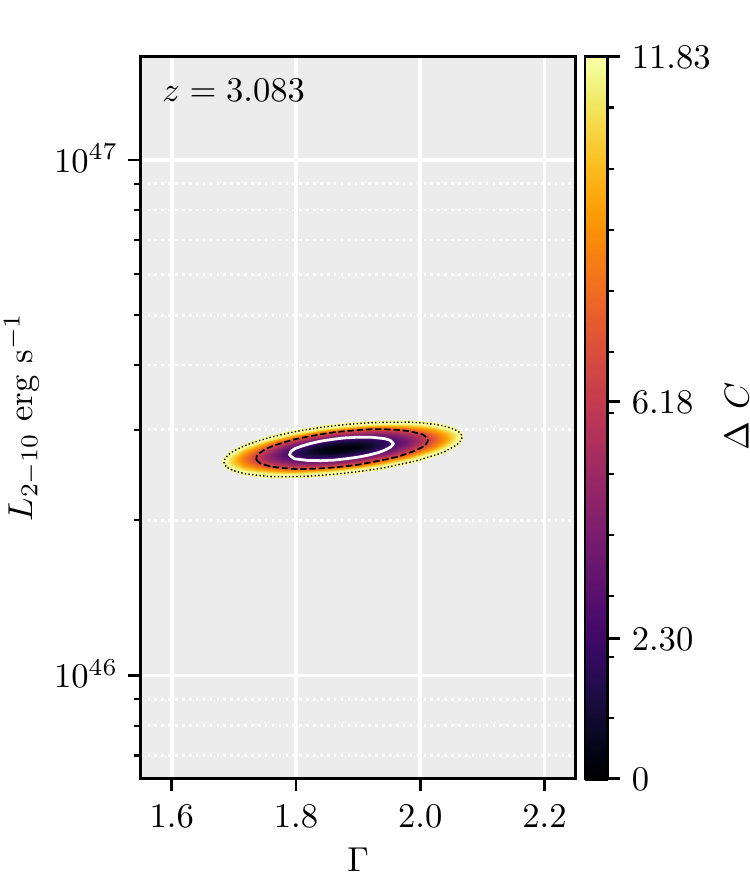}
   \includegraphics{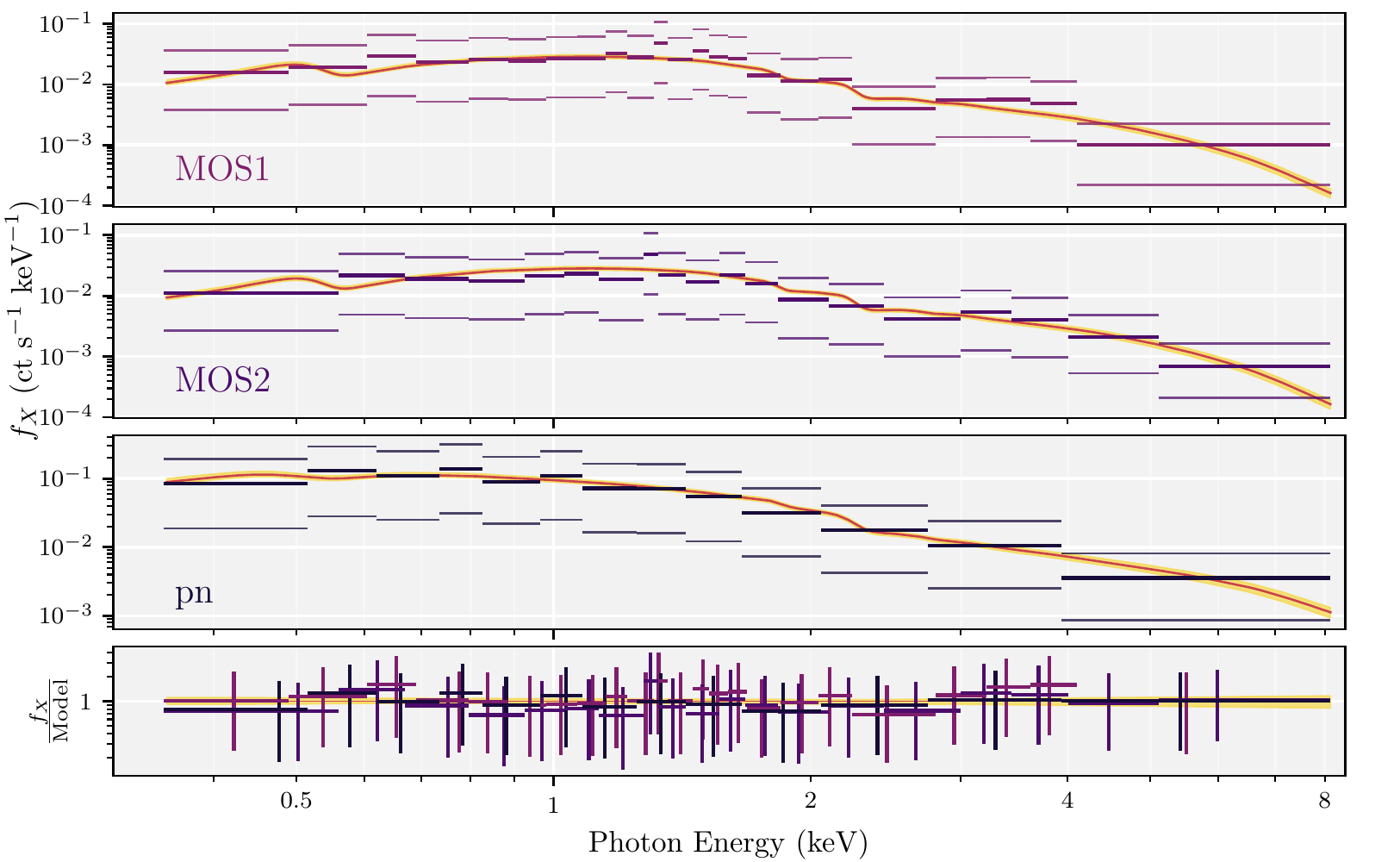}
      \caption{\small EPIC observations of GraL J0659+1629. The description of the panels is given in the text.
      }\label{fig:apx0659}\label{fig:first_appx}
      \vspace{-5mm}
\end{figure*}

To assist in the planning of future observations of these quasars should they be the site of a future microlensing event, we present full details of our fits and analysis in Figures \ref{fig:first_appx} and \ref{fig:apx_upperlimits}. In the upper left panels, we show the individual images from the three EPIC cameras, as well as a combined view of all three. These images are $130^{\prime\prime}$ on a side, are centered on the position of the quasar, and are smoothed with Gaussian kernels of width $4\farcs0$. In the upper right, we show contours of the best-fitting values of $\Gamma$ and $L_X$. Contours corresponding to $1$, $2$, and $3\sigma$ are indicated by the white, black dashed, and black dotted lines, respectively. For the two quasars well-fit by including a redshifted absorption component, these contours trace the lowest value of $\Delta C$ for a given pair of $\Gamma$ and luminosity across the entire range of modeled column densities.

In the bottom panels of Figure \ref{fig:first_appx}, we show the individual observed and best-fitting spectra in the three EPIC cameras, as well as the residuals. Observed spectra are background-subtracted and have been binned for presentation purposes. Thin horizontal lines above and below the observed values correspond to $1\sigma$ uncertainties. Best fit models, propogated through \texttt{fakeit} in \texttt{XSPEC}, are shown by the red lines, while yellow regions trace the bounds of $1\sigma$ uncertainties on the fit. For the pn observations that were unusable, the subfigure is rendered in gray.

\renewcommand{\thefigure}{\arabic{figure} (Cont.)}

\addtocounter{figure}{-1}
\begin{figure*}
\centering
\vspace{5mm}
   \includegraphics{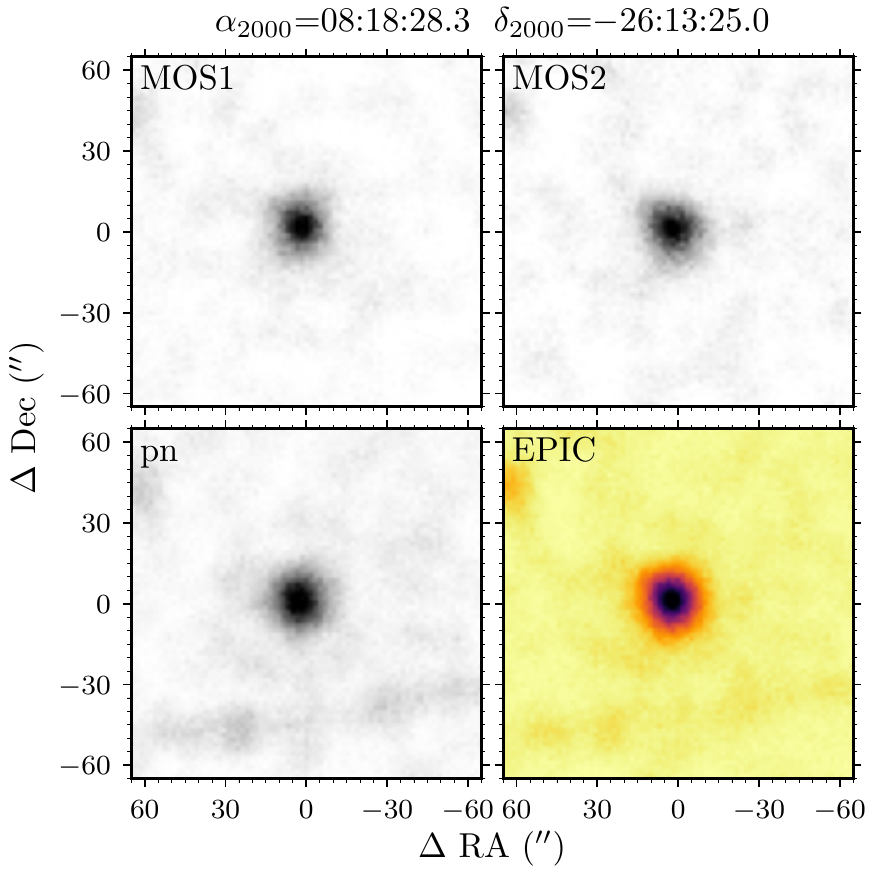}
   \includegraphics{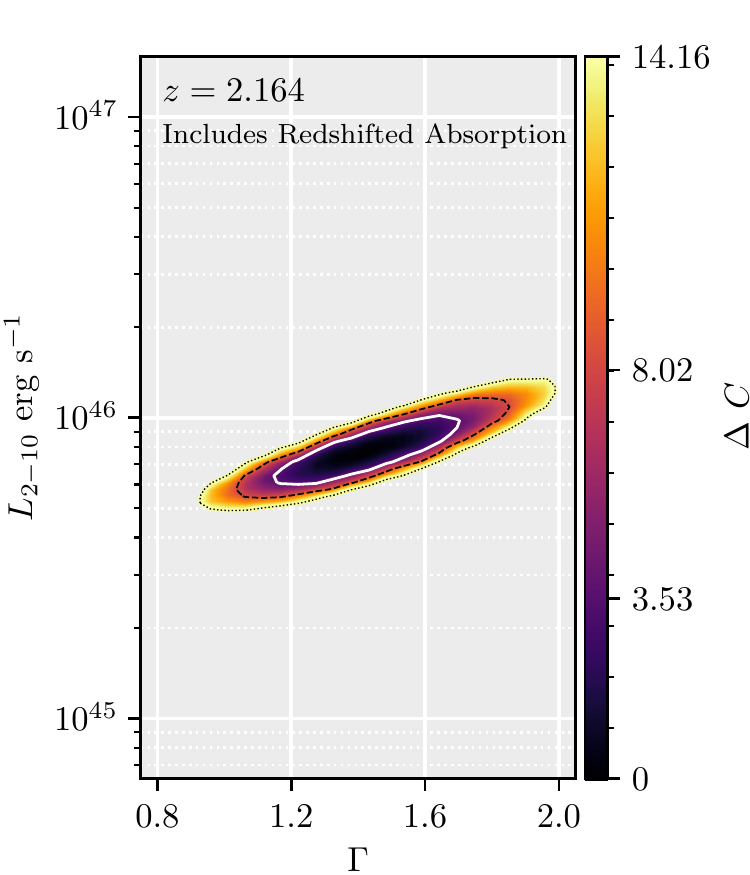}
   \includegraphics{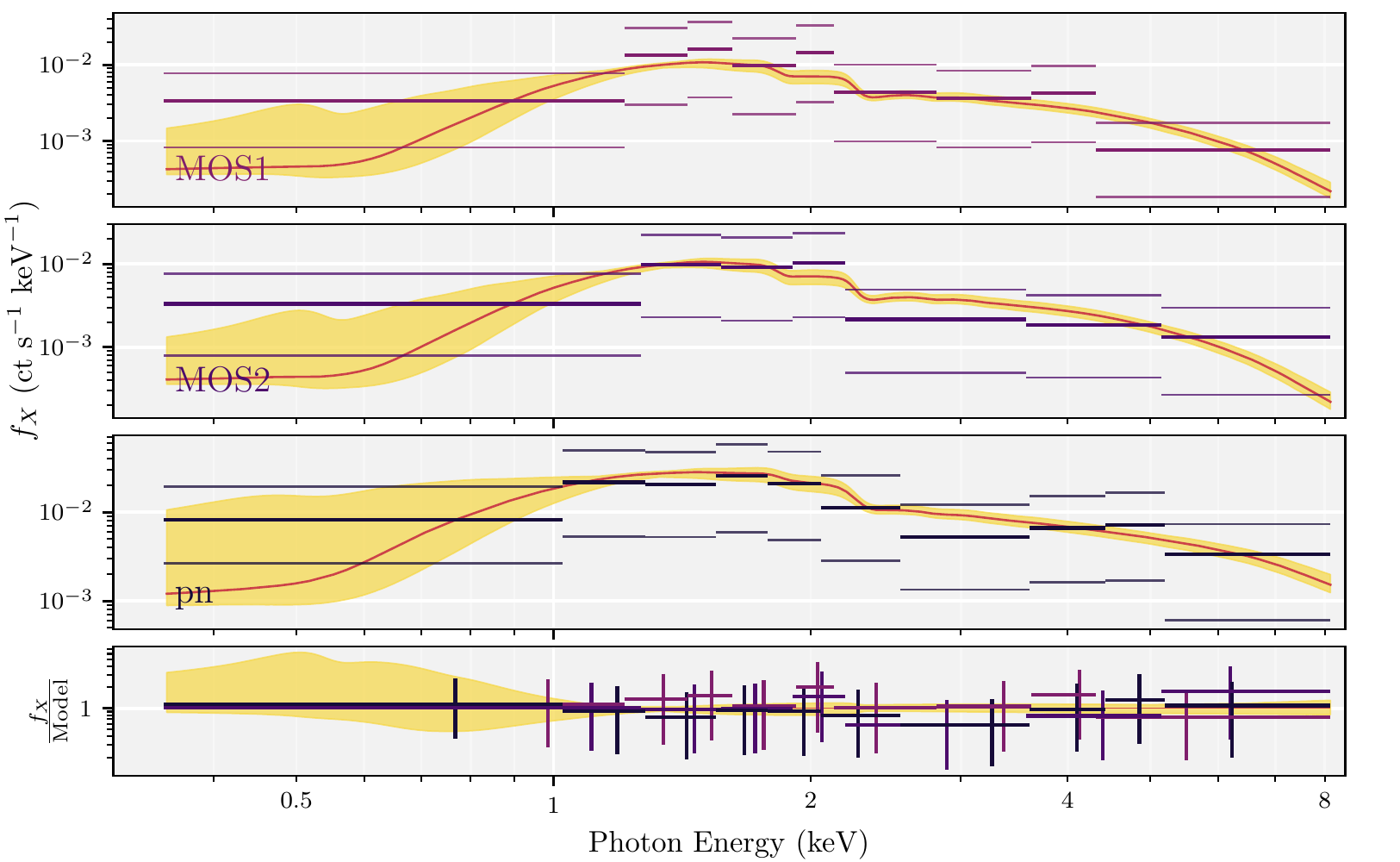}
      \caption{\small EPIC observations of GraL J0818${-}$2613. This fit includes a redshifted absorption component, and so the contours presented here show the best fitting value for a given $\Gamma$-$L_X$ pair at any obscuration. $\Delta C$ contours are spaced to account for three parameters of interest.
      }\label{fig:apx0818}
      \vspace{-5mm}
\end{figure*}

\addtocounter{figure}{-1}
\begin{figure*}
\centering
\vspace{5mm}
   \includegraphics{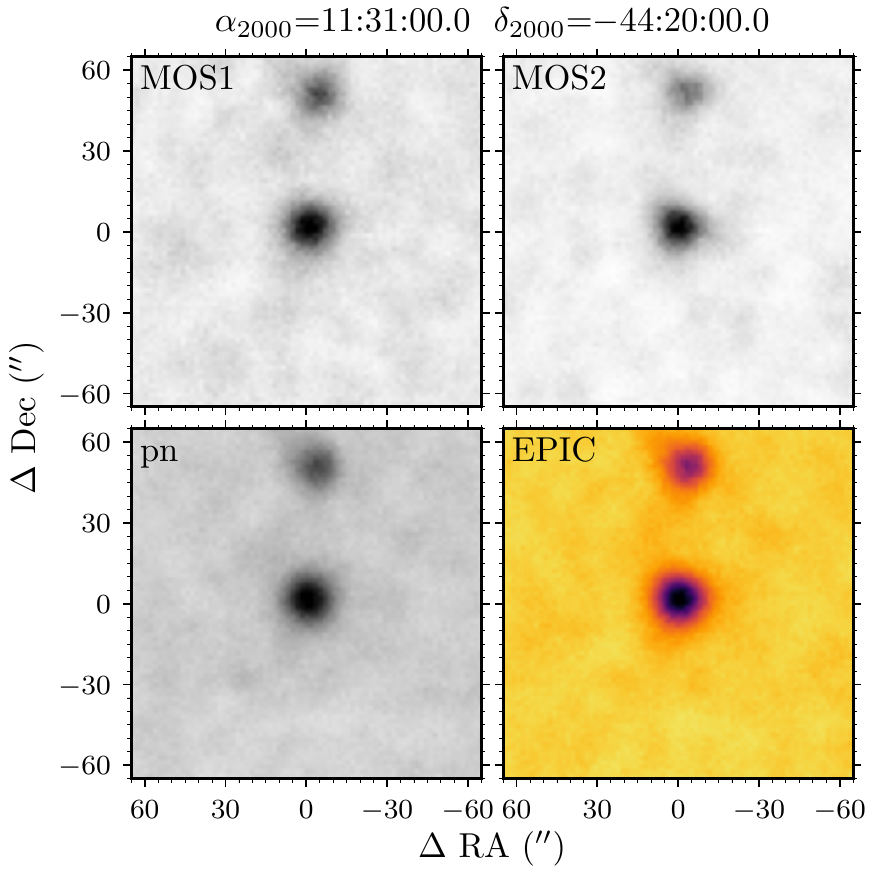}
   \includegraphics{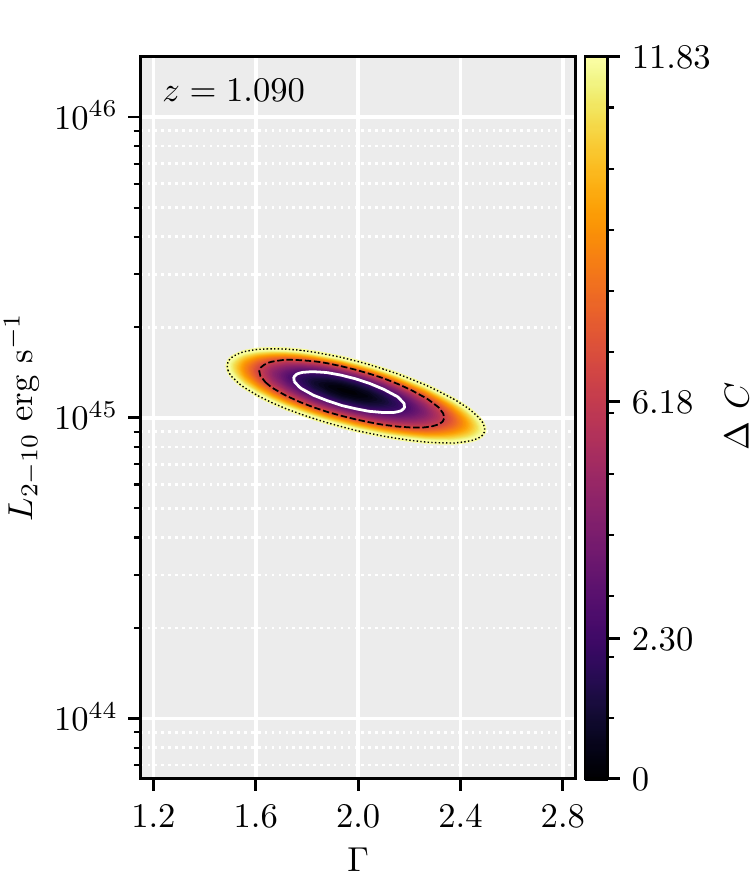}
   \includegraphics{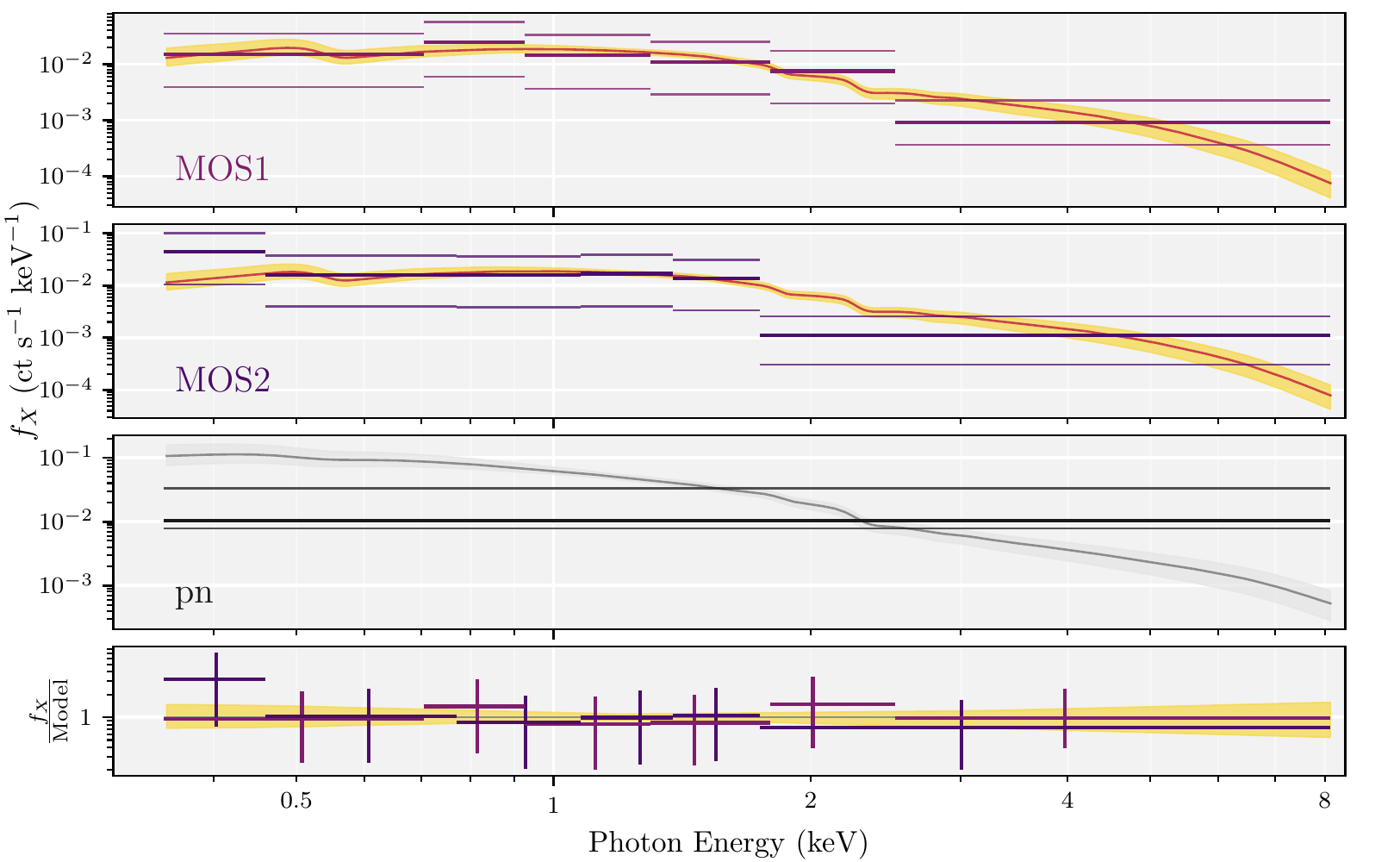}
      \caption{\small EPIC observations of GraL J1131$-$4419. The pn camera was not included in the spectral fitting of this source. A second, fainter source is visible ${\sim}55^{\prime\prime}$ to the north, as discussed in \S~\ref{ssec:individuals}.
      }\label{fig:apx1131}
      \vspace{-5mm}
\end{figure*}

\addtocounter{figure}{-1}
\begin{figure*}
\centering
\vspace{5mm}
   \includegraphics{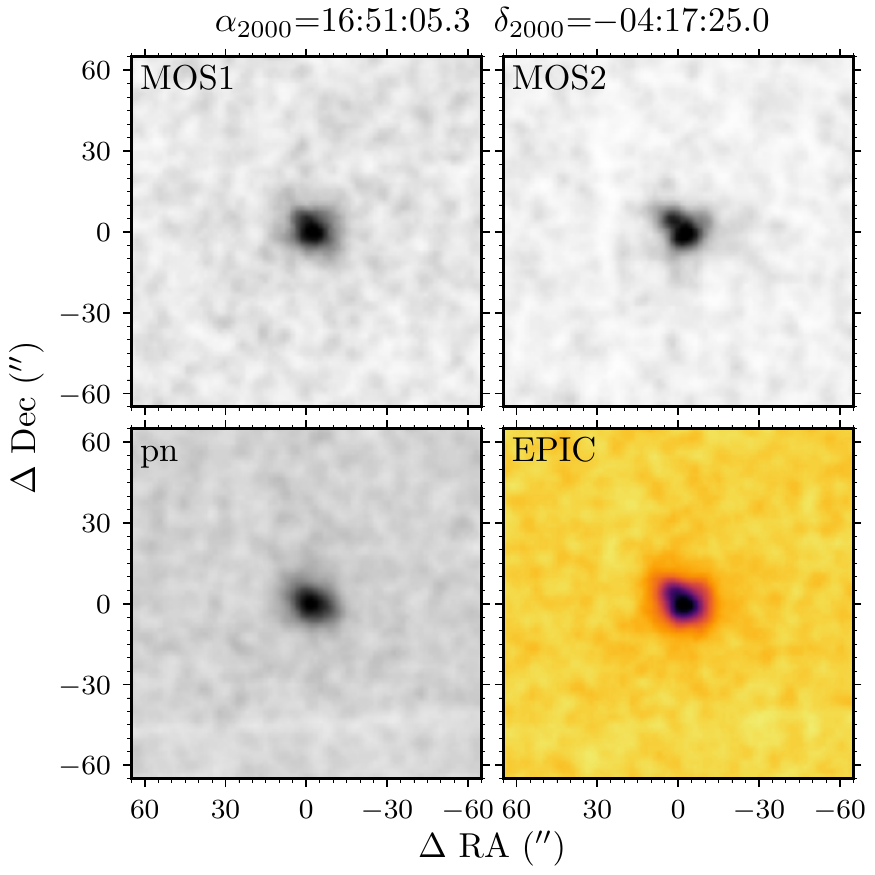}
   \includegraphics{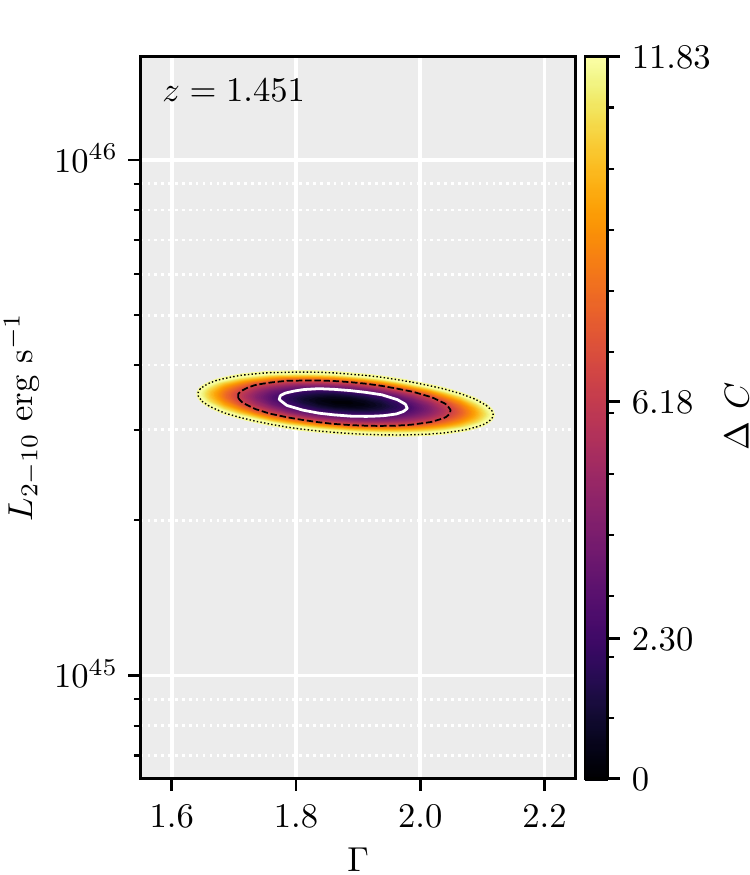}
   \includegraphics{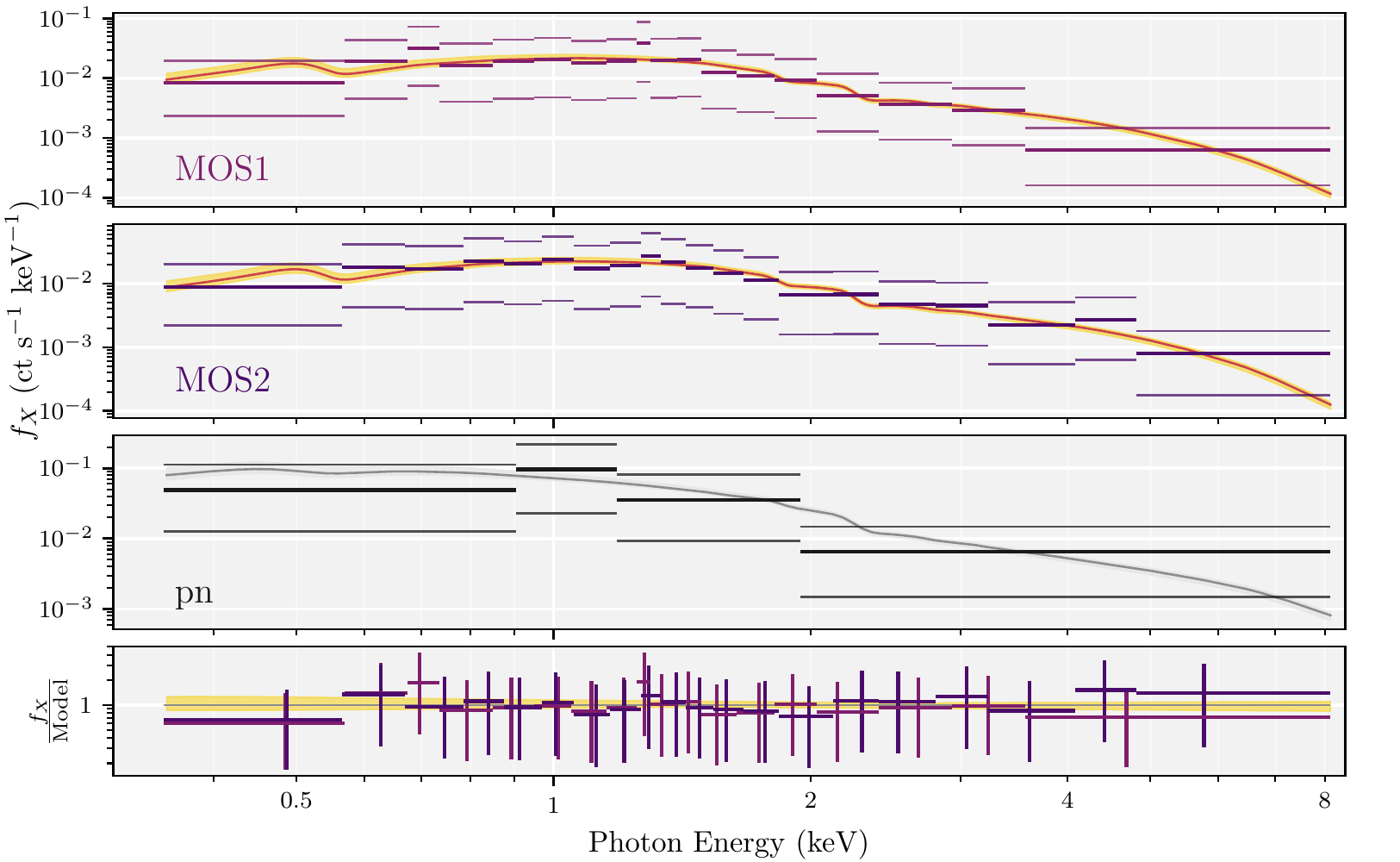}
      \caption{\small EPIC observations of GraL J1651$-$0417. The smoothing scale has been reduced to a width of $2\farcs0$ for this image to highlight that the large separation is resolved by these observations. The pn camera was not included in the spectral fitting of this source.
      }\label{fig:apx1651}
      \vspace{-5mm}
\end{figure*}

\addtocounter{figure}{-1}
\begin{figure*}
\centering
\vspace{5mm}
   \includegraphics{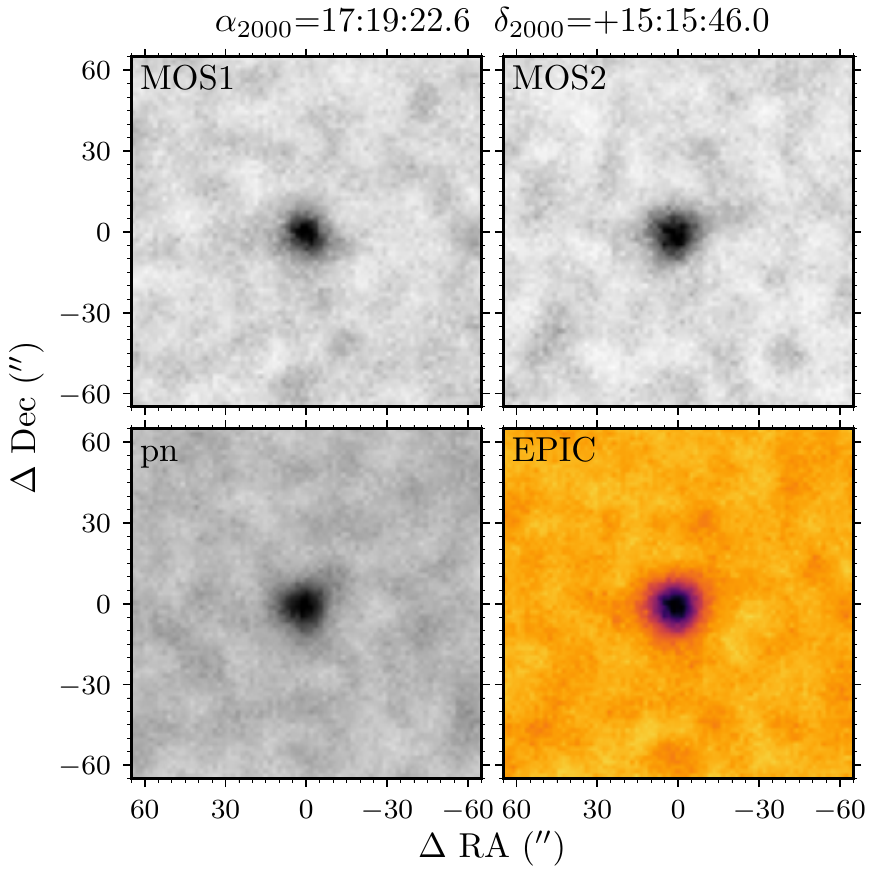}
   \includegraphics{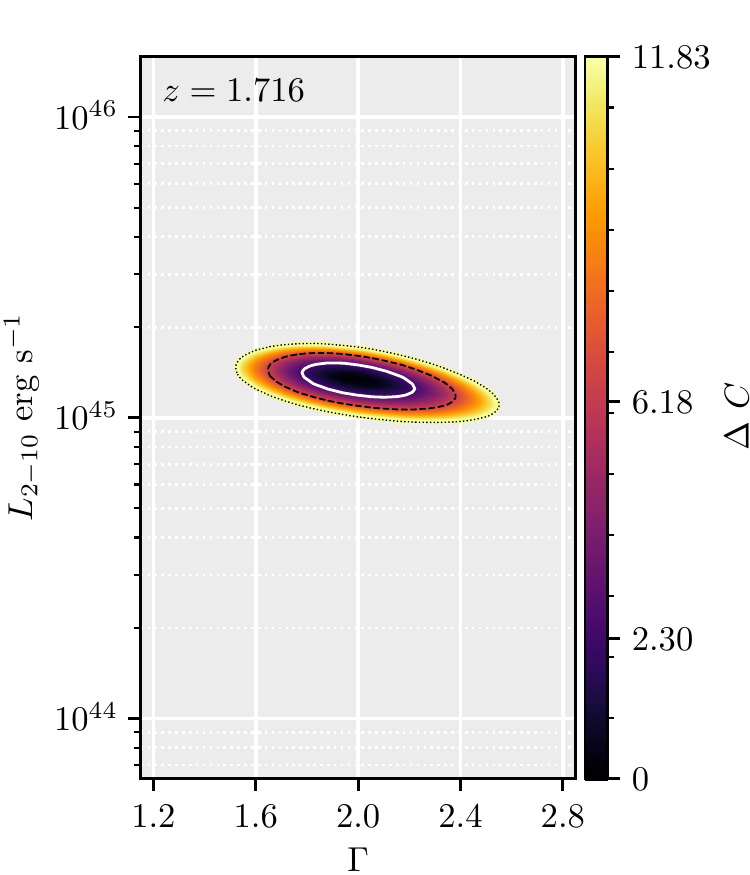}
   \includegraphics{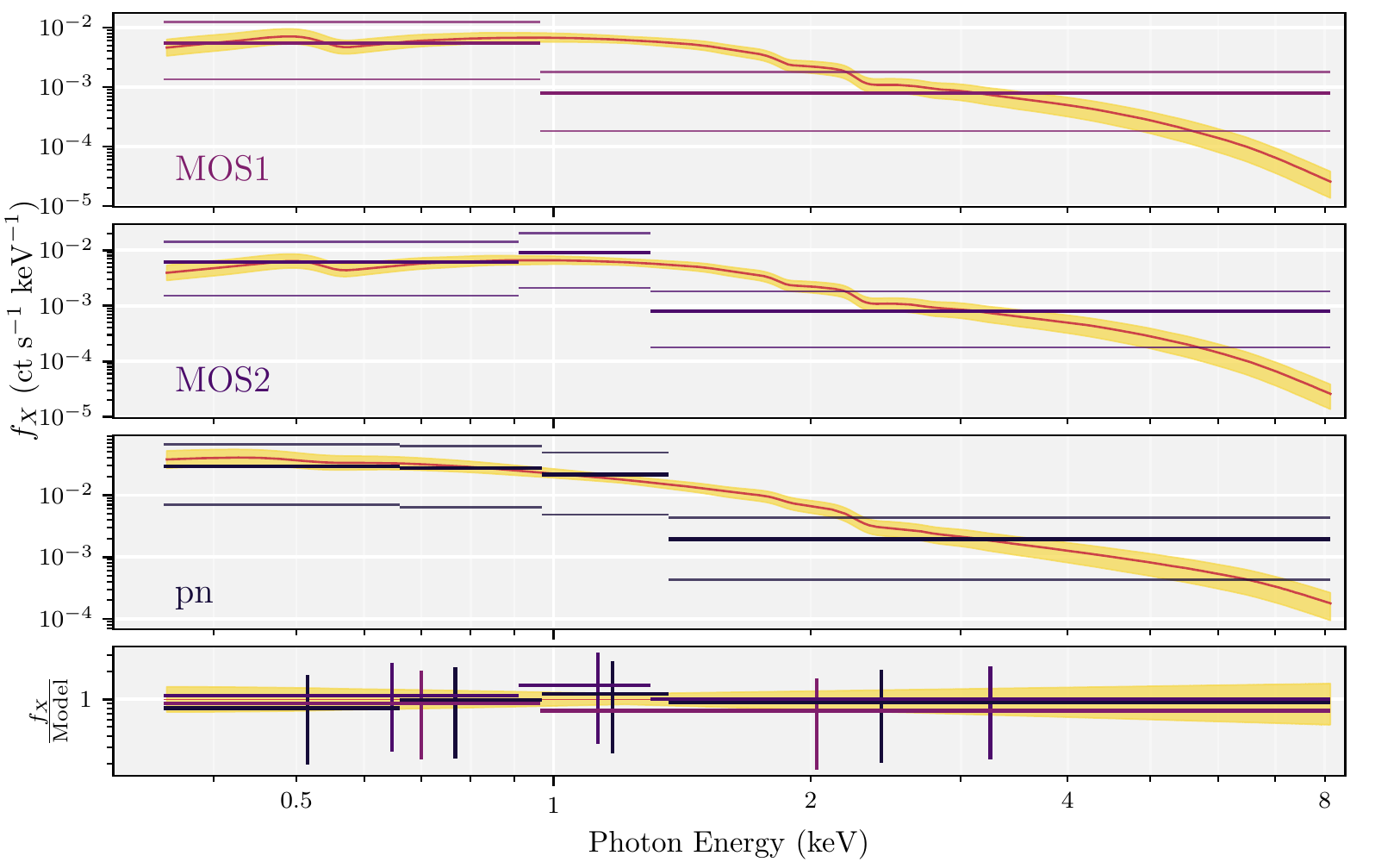}
      \caption{\small EPIC observations of GraL J1719+1515.
      }\label{fig:apx1719}
      \vspace{-5mm}
\end{figure*}

\addtocounter{figure}{-1}
\begin{figure*}
\centering
\vspace{5mm}
   \includegraphics{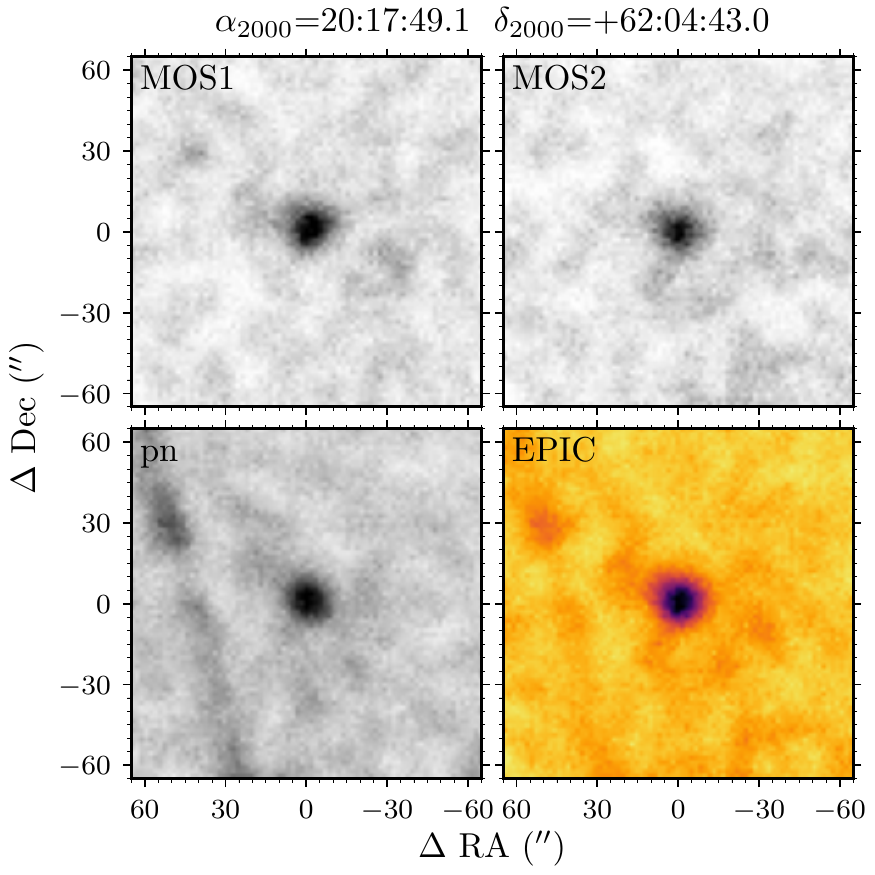}
   \includegraphics{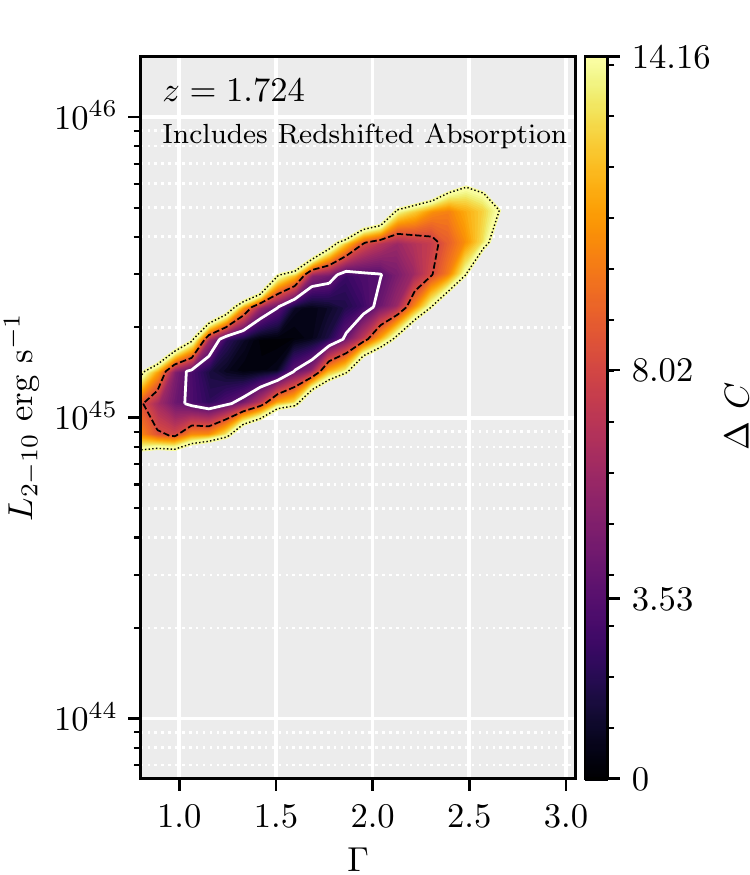}
   \includegraphics{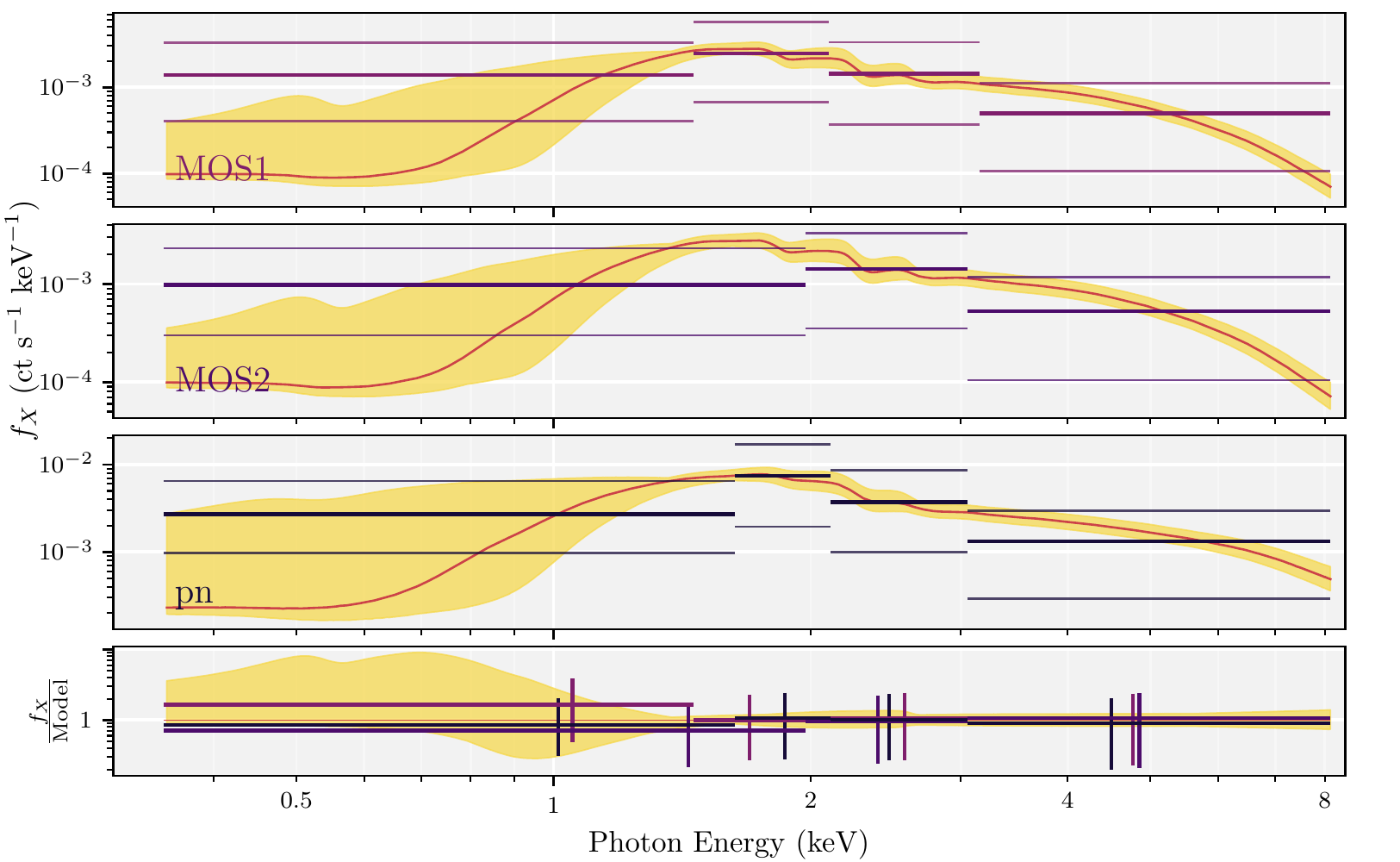}
      \caption{\small EPIC observations of GraL J2017+6204. This fit includes a redshifted absorption component, and so the contours presented here show the best fitting value for a given $\Gamma$-$L_X$ pair at any obscuration. $\Delta C$ contours are spaced to account for three parameters of interest.
      }\label{fig:apx2017}
      \vspace{-5mm}
\end{figure*}

\addtocounter{figure}{-1}
\begin{figure*}
\centering
\vspace{5mm}
   \includegraphics{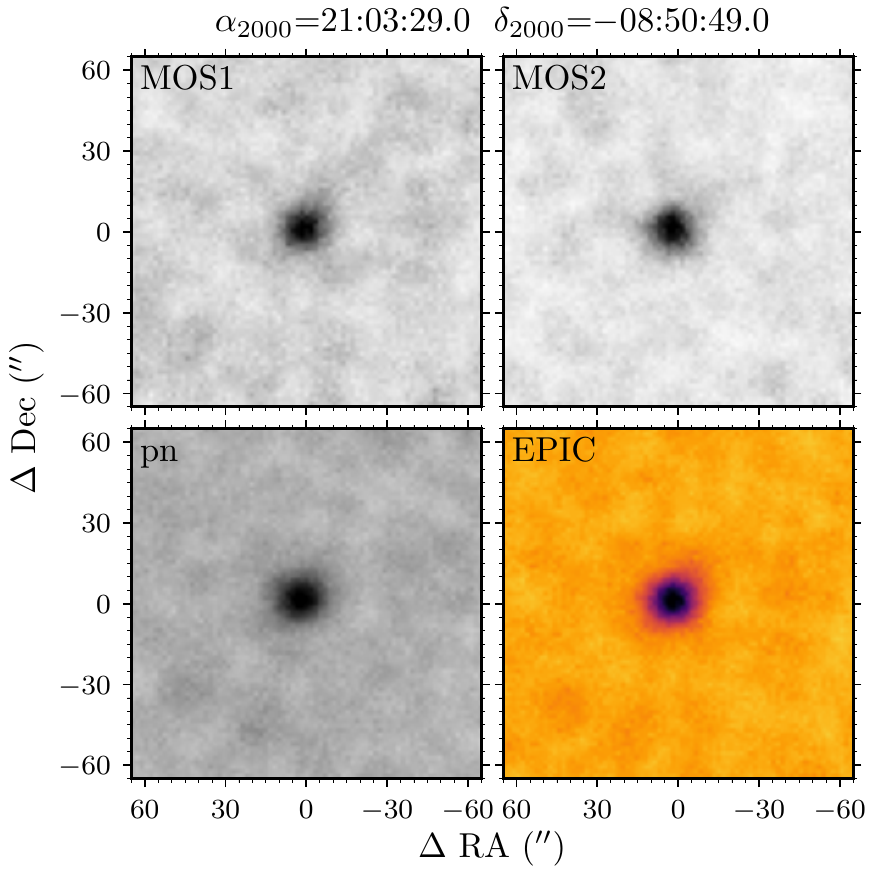}
   \includegraphics{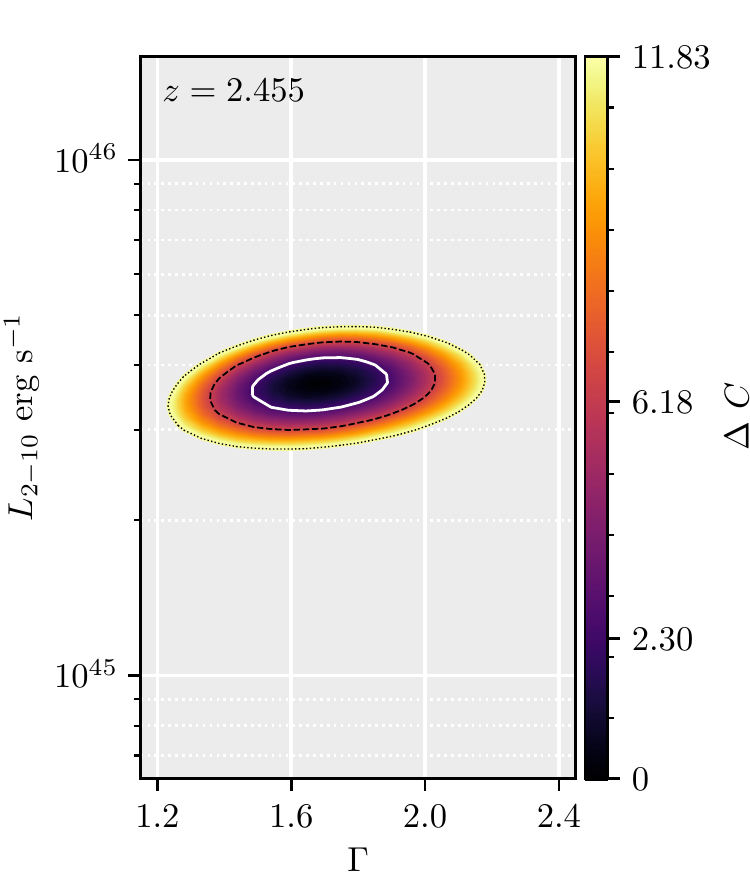}
   \includegraphics{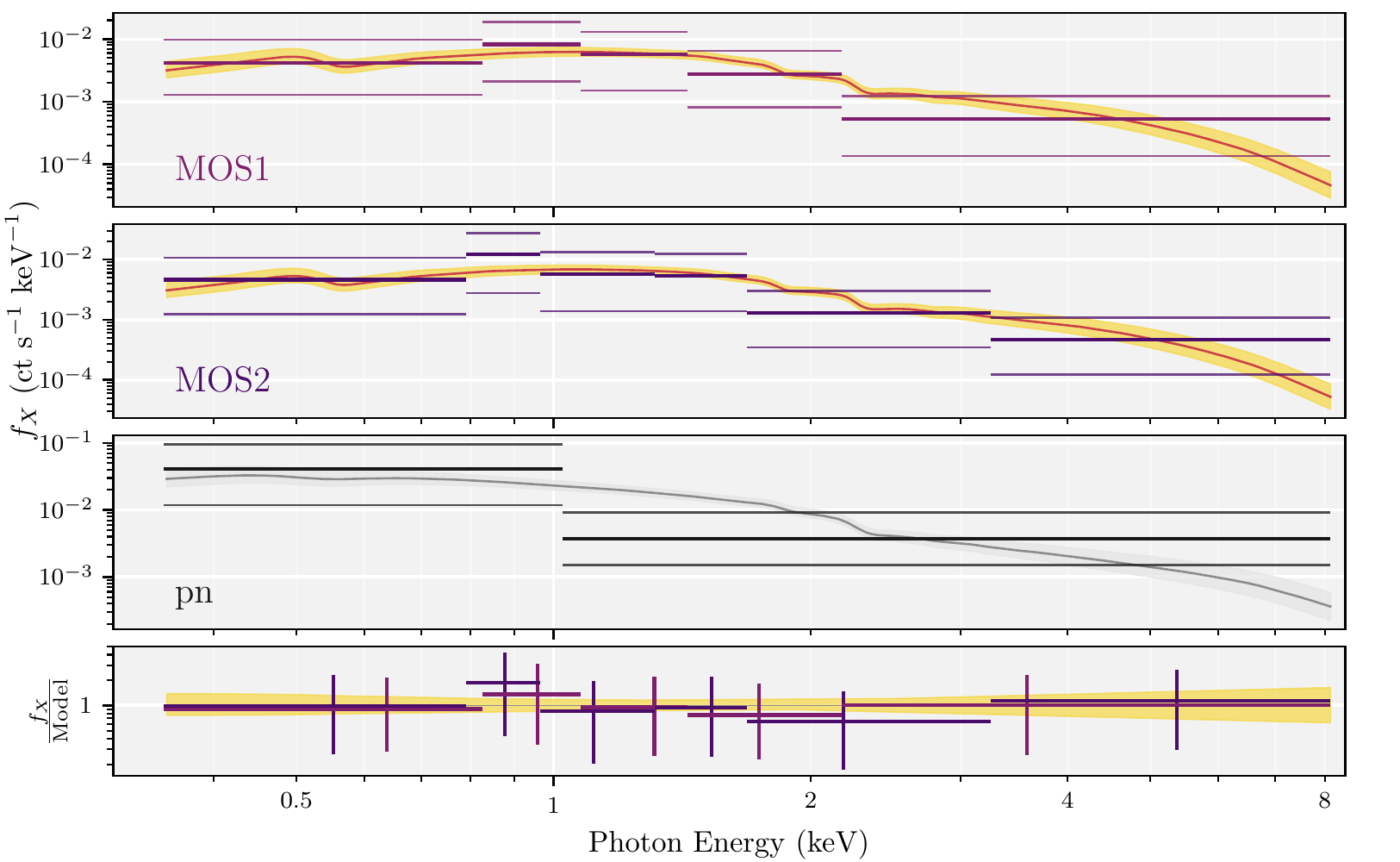}
      \caption{\small EPIC observations of GraL J2103$-$0850. The pn camera was not included in the spectral fitting of this source.
      }\label{fig:apx2103}
      \vspace{-5mm}
\end{figure*}

\addtocounter{figure}{-1}
\begin{figure*}
\centering
\vspace{5mm}
   \includegraphics{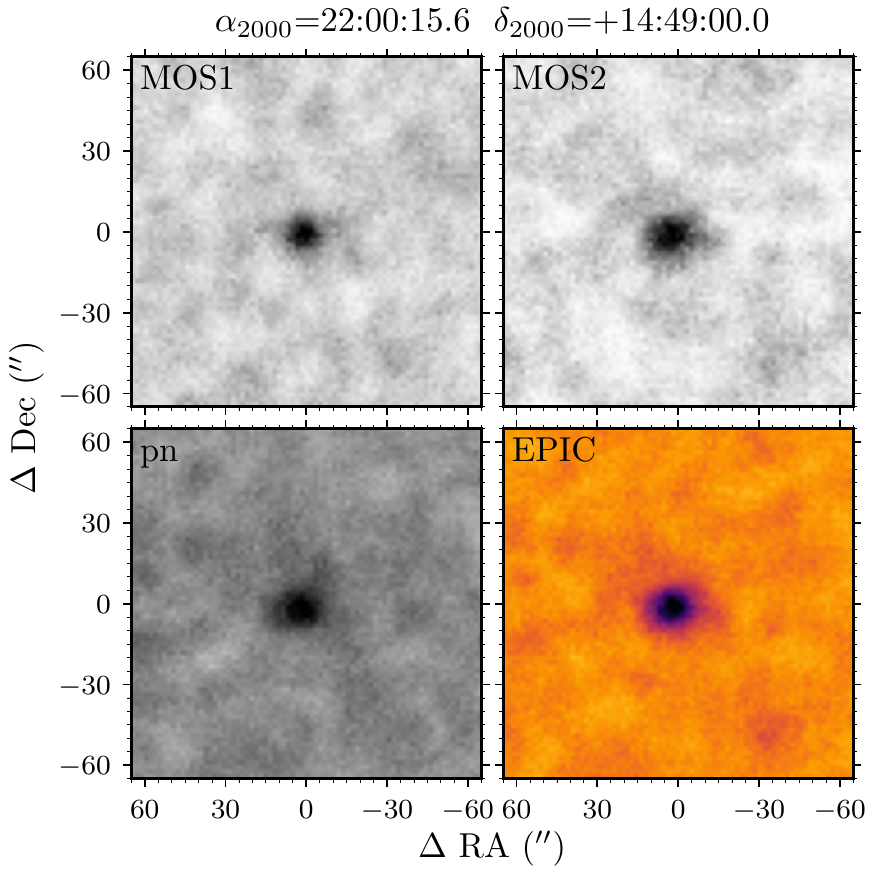}
   \includegraphics{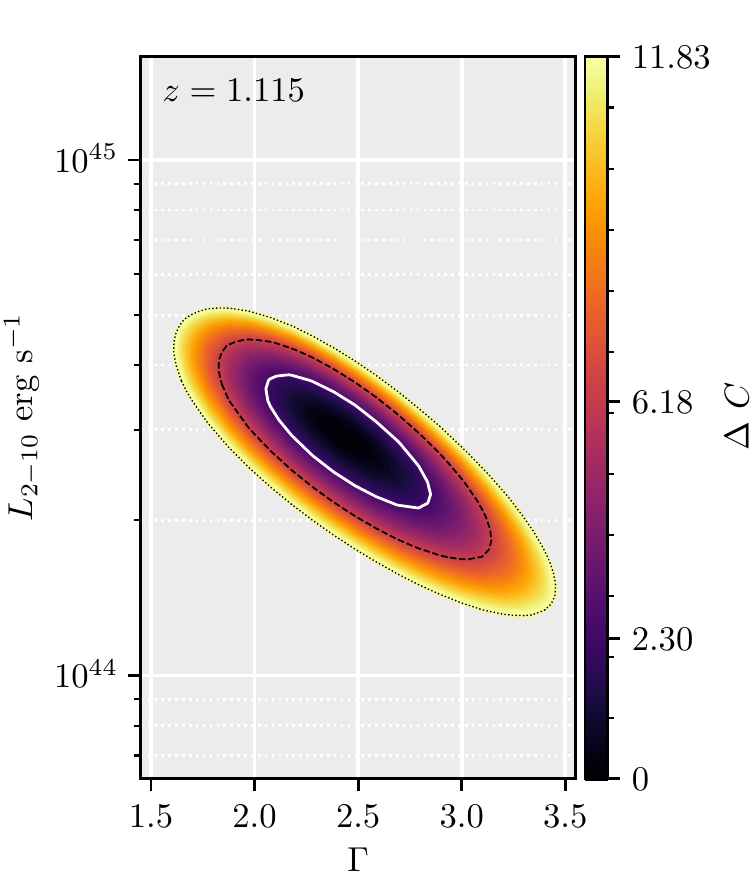}
   \includegraphics{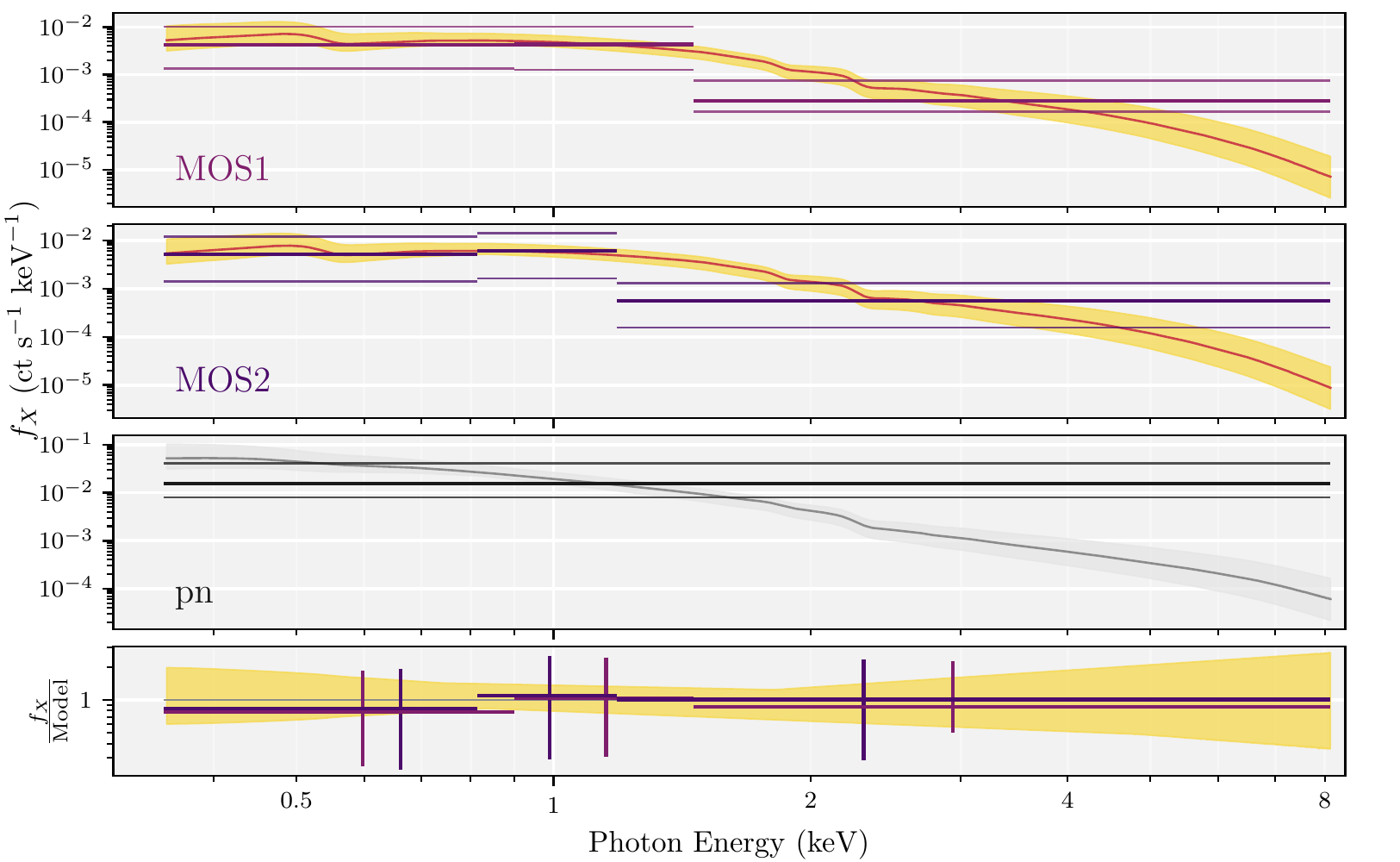}
      \caption{\small EPIC observations of GraL J2200+1448. The pn camera was not included in the spectral fitting of this source.
      }\label{fig:apx2200}
      \vspace{-5mm}
\end{figure*}

\renewcommand{\thefigure}{\arabic{figure}}

\begin{figure*}
\centering
\vspace{5mm}
   \includegraphics{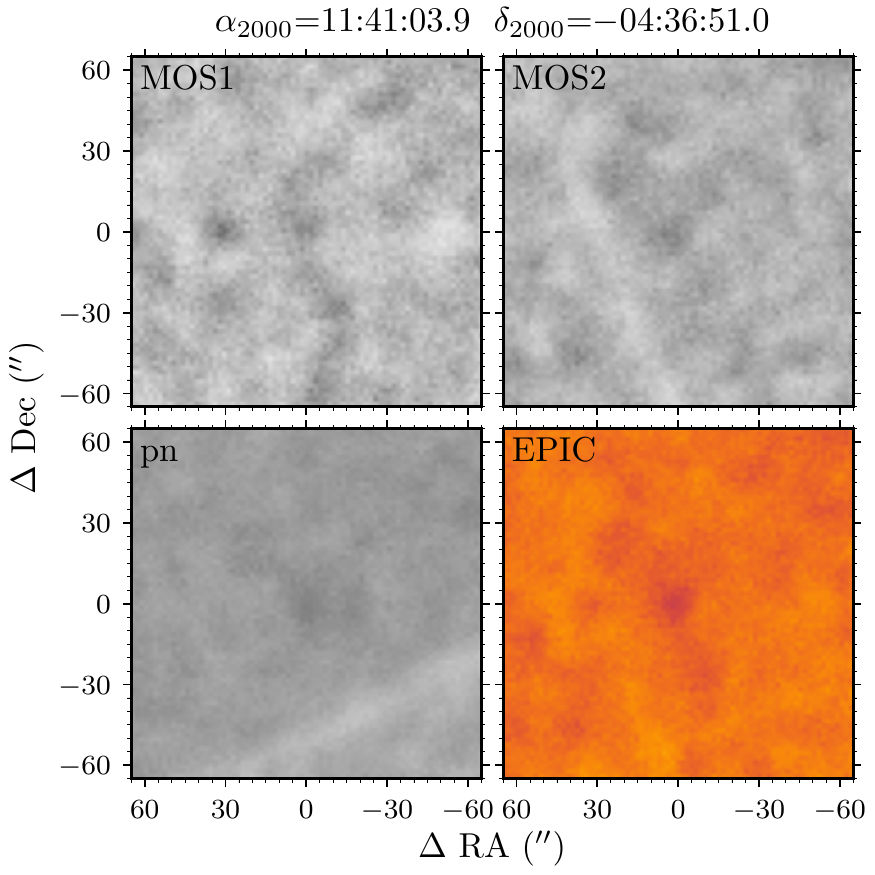}
   \includegraphics{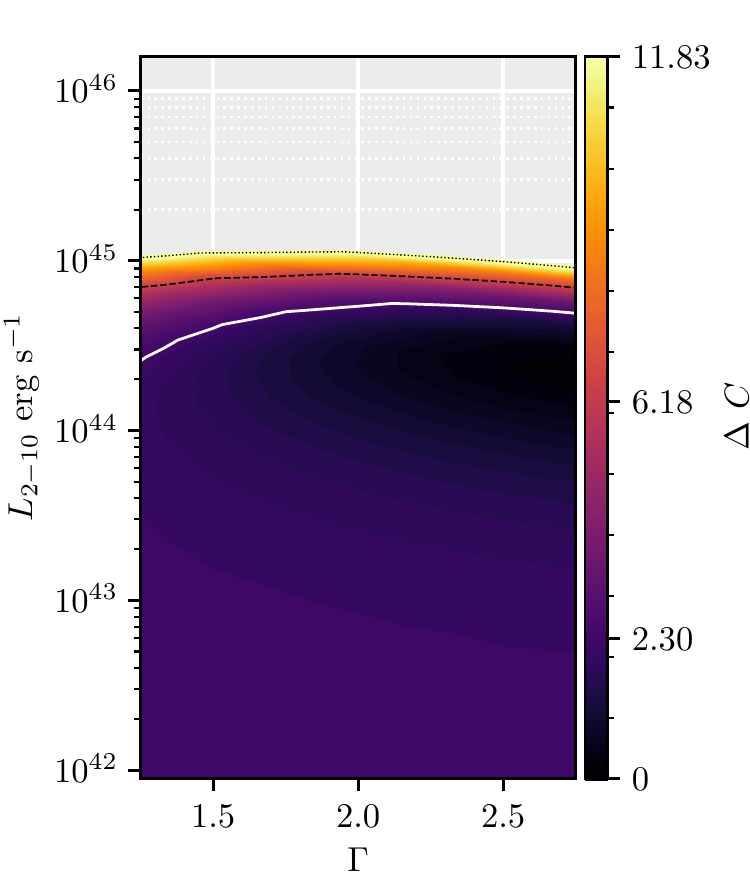}
   \includegraphics{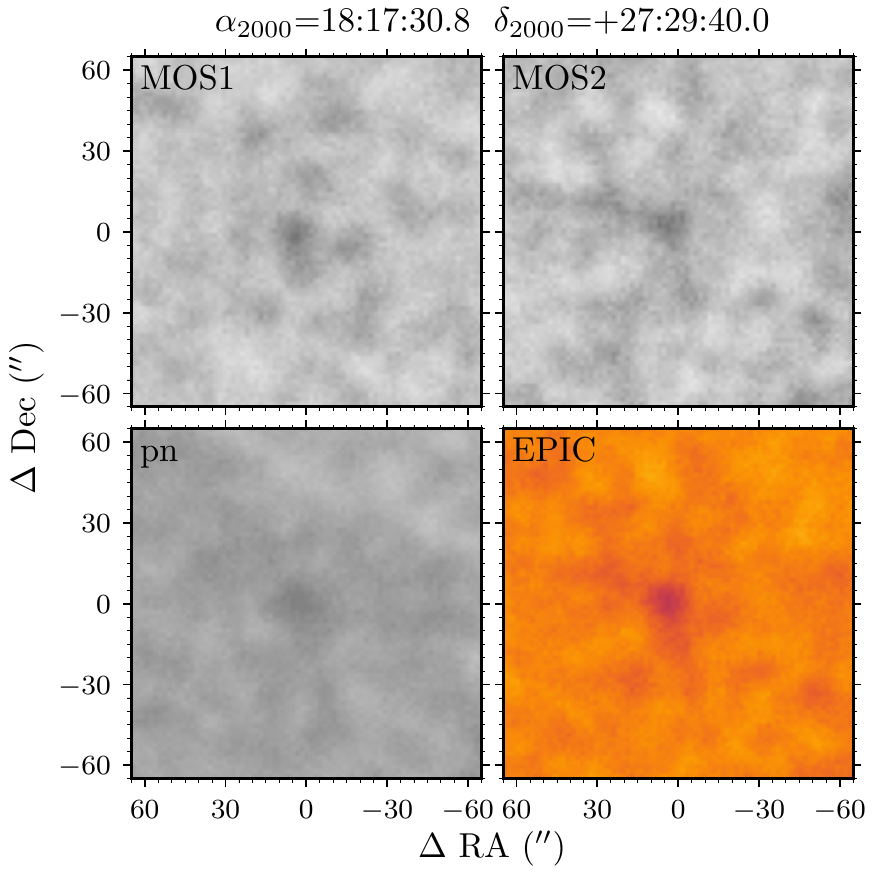}
   \includegraphics{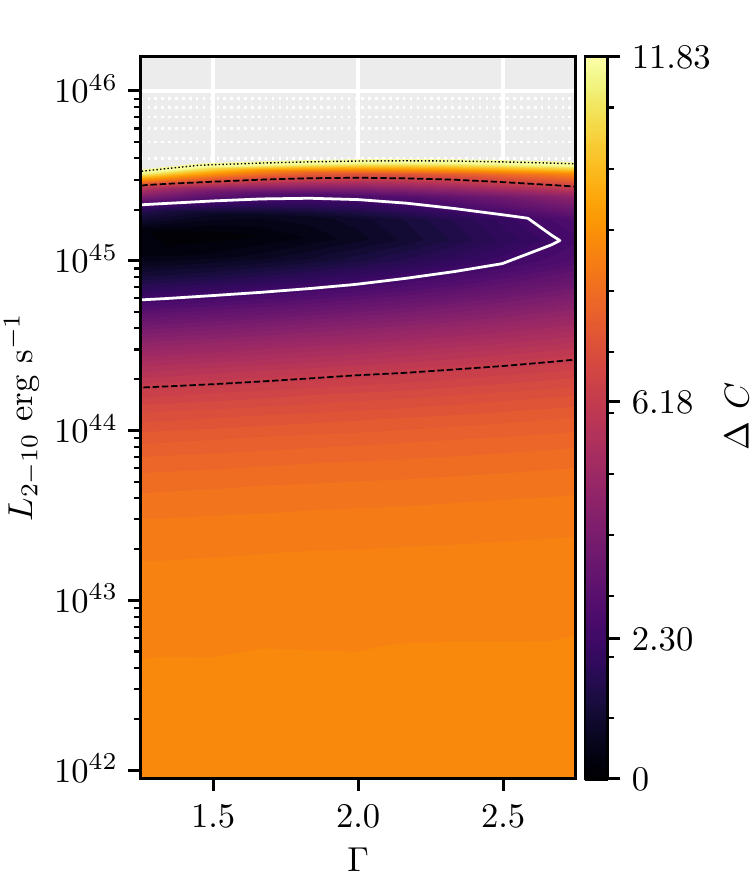}
      \caption{\small EPIC observations of SDSS J1141${-}$0436 ({\bf top}) and GraL J1817${+}$2729 ({\bf bottom}), in the same format as Figure \ref{fig:first_appx}. As these objects are not strongly detected, spectral fits are not shown, only the contours of upper limits on their X-ray luminosities.
      }\label{fig:apx_upperlimits}
      \vspace{-5mm}
\end{figure*}

\allauthors
\end{document}